\documentclass[aps,pra,10pt,twocolumn,superscriptaddress]{revtex4-1}
\usepackage{amsmath}
\usepackage{amsfonts}
\usepackage{amssymb}
\usepackage{graphicx}
\usepackage{color}
\usepackage{hyperref}
\hypersetup{colorlinks=true,allcolors=blue}

\newcommand{\OP}[2]{\vert #1\rangle\langle #2\vert}
\newcommand{\BRA}[1]{\langle #1\vert}
\newcommand{\KET}[1]{\vert #1\rangle}
\newcommand{\ABS}[1]{\vert #1\vert}
\newcommand{\ddt}{\frac{\mathrm{d}}{\mathrm{d}t}}
\newcommand{\LB}[2]{\mathcal{L}_{#1,#2}}
\newcommand{\Tr}[1]{\mathrm{Tr}\left\lbrace #1 \right\rbrace}
\newcommand{\aH}{\hat{a}_{\text{H}}}
\newcommand{\aHd}{\aH^\dagger}
\newcommand{\aV}{\hat{a}_{\text{V}}}
\newcommand{\aVd}{\aV^\dagger}
\newcommand{\dcL}[1]{\Delta_{\text{#1}}}
\newcommand{\aD}{\hat{a}_{\text{D}}}
\newcommand{\aDd}{\aD^\dagger}
\newcommand{\aA}{\hat{a}_{\text{A}}}
\newcommand{\aAd}{\aA^\dagger}

\begin{document}

\title{Different types of photon entanglement from a constantly driven quantum emitter inside a cavity}

\author{T. Seidelmann}
\affiliation{Lehrstuhl f{\"u}r Theoretische Physik III, Universit{\"a}t Bayreuth, 95440 Bayreuth, Germany}
\author{M. Cosacchi}
\affiliation{Lehrstuhl f{\"u}r Theoretische Physik III, Universit{\"a}t Bayreuth, 95440 Bayreuth, Germany}
\author{M. Cygorek}
\affiliation{Heriot-Watt University, Edinburgh EH14 4AS, United Kingdom}
\author{D. E. Reiter}
\affiliation{Institut f{\"u}r Festk{\"o}rpertheorie, Universit{\"a}t M{\"u}nster, 48149 M{\"u}nster, Germany}
\author{A. Vagov}
\affiliation{Lehrstuhl f{\"u}r Theoretische Physik III, Universit{\"a}t Bayreuth, 95440 Bayreuth, Germany}
\affiliation{ITMO University, St. Petersburg, 197101, Russia}
\author{V. M. Axt}
\affiliation{Lehrstuhl f{\"u}r Theoretische Physik III, Universit{\"a}t Bayreuth, 95440 Bayreuth, Germany}

\begin{abstract}
Bell states are the most prominent maximally entangled photon states. In a typical four-level emitter, like a semiconductor quantum dot, the photon states exhibit only one type of Bell state entanglement. By adding an external driving to the emitter system, also other types of Bell state entanglement are reachable without changing the polarization basis. In this paper, we show under which conditions the different types of entanglement occur and give analytical equations to explain these findings. We further identify special points, where the concurrence, being a measure for the degree of entanglement, drops to zero, while the coherences between the two-photon states stay strong. Results of this work pave the way to achieve a controlled manipulation of the entanglement type in practical devices. 
\end{abstract}

\maketitle

\section{Introduction}
\label{sec:introduction}

Entanglement of quantum states is one of the most remarkable and interesting physical effects that separate the quantum mechanical from the classical world \cite{Horodecki:09,Orieux_entangled}. Entanglement can be used to test quantum mechanical principles on a fundamental level, e.g., by revealing violations of Bell inequalities \cite{entangled-photon2,Orieux_entangled}. Furthermore, many fascinating and innovative applications, e.g., in quantum cryptography \cite{Gisin:02,Lo_quantum_cryptography}, quantum communication \cite{duan_quantum_comm,Huber_overview_2018}, or quantum information processing and computing \cite{pan:12,Bennett:00,Kuhn:16,Zeilinger_entangled}, rely on entangled photon pairs.

The defining property of an entangled bipartite system is that its quantum mechanical state cannot be factorized into parts corresponding to the constituent subsystems. There are four prominent states, which are maximally entangled and known as the Bell states, established for two entangled photons with horizontal $H$ polarization and vertical $V$ polarization
\begin{eqnarray}
\label{eq:def_pol_ent_state}
\KET{\Phi_\pm} = \frac{1}{\sqrt{2}}\left( \KET{HH} \pm \KET{VV} \right) ,\\
\label{eq:def_bell_ent_state}
\KET{\Psi_\pm} = \frac{1}{\sqrt{2}}\left( \KET{HV} \pm \KET{VH} \right) \,.
\end{eqnarray}
In the following we will refer to these states as $\Phi$ Bell state ($\Phi$BS) and  $\Psi$ Bell state ($\Psi$BS). To create maximally entangled states, one of the best established routes is via the cascaded relaxation in few-level systems like atoms, F-centers or semiconductor quantum dots \cite{edamatsu2007entangled}. 

In this paper, we study under which driving conditions, a four-level emitter (FLE) placed in a microcavity produces entangled photons being either in a $\Phi$BS or $\Psi$BS. We demonstrate that a constantly driven FLE undergoes a sharp transition between regions of high $\Phi$BS and $\Psi$BS entanglement for a certain two-photon resonance. At the transition the degree of entanglement drops to zero at a special point, because the quantum state of the system becomes factorizable. We will further study all two-photon resonances revealing a rich variety of different scenarios with or without switching the type of entanglement and with or without special points of zero concurrence.

\section{Generation of entangled states}
\label{sec:generation_entanglement}
The generation procedure of entangled photons in a typical (non-driven) four-level system is as follows [see also Fig.~\ref{fig:model}(left)]: In a first step the uppermost state is prepared, e.g., by using two-photon resonant or near-resonant excitation with short coherent pulses \cite{entangled-photon1,hanschke2018,huber2017,reindl2017,Finley_phonon-assisted,PI_phonon-assisted_biexc_prep-exp,PI_phonon-assisted_biexc_prep,PI_undressing,Reiter_2014} or adiabatic rapid passage protocols \cite{Debnath_ARP_biexciton,Glaessl_ARP_biexciton,Kaldewey_ARP_biexciton,Uebersichtsartikel_2019}. The excited emitter then decays into one of two different intermediate states emitting either a horizontally or vertically polarized photon. In the subsequent decay to the ground state a second photon is emitted, which has the same polarization as the first one. In an ideal situation there is no which-path information and the resulting two-photon state is a $\Phi$BS. Experiments and theoretical studies in semiconductor quantum dots demonstrated the possibility to generate $\Phi$BS entanglement \cite{Seidelmann2019,Different-Concurrences:18, Phon_enhanced_entanglement,Jahnke2012, heinze17,BiexcCasc_Carmele,Stevenson2006,Young_2006,Muller_2009,Huber_PRL_2018, Wang_2019,Liu2019, Bounouar18,dousse:10,winik:2017, entangled-photon1,Fognini_2019,entangled-photon1, entangled-photon2,Hafenbrak, Biexc_FSS_electrical_control_Bennett, EdV,Troiani2006,stevenson:2012, Benson_2000_QD_cav_device, winik:2017,Huber_PRL_2018,Wang_2019, Liu2019,EdV}.

The situation changes profoundly when the few-level system is continuously driven by an external laser. Then additionally, it become possible to create $\Psi$BS entanglement. A possible mechanism could be that the uppermost state emits a horizontally polarized photon via one path way, is then re-excited by the laser and then emits a vertically polarized photon via the other path. Since the sequence of emission of a pair of $H,V$ or $V,H$ polarized photons is identical, this process results in an entangled $\Psi$BS.
Note that the states $\KET{HV}$ and $\KET{VH}$ are distinguished by the temporal order of the $H$ or $V$ polarized photon emissions.
Indeed, Mu{\~n}oz \textit{et al.} \cite{munoz15} found that under specific conditions the resulting two-photon state is close to the $\Psi$BS. Here we will show that $\Psi$BS entanglement occurs under various conditions, but also $\Phi$BS entanglement is supported by a driven FLE system. 

To create entangled photon states in an optimal way, the FLE is embedded inside a microcavity. By this, the coupling to the cavity enhances the light-collection efficiency and the photon emission rate due to the Purcell effect \cite{dousse:10,Badolato:05}. Additionally, the energetic placement of the cavity modes can have a profound impact on the resulting degree of entanglement. By placing the cavity modes in resonance with a two-photon transition of the emitter \cite{EdV,Jahnke2012,heinze17,Seidelmann2019, Ota_2pht_emis_cavity_2011,munoz15} direct two-photon emission processes dominate over sequential single-photon ones. Since the direct two-photon emission is much less affected by a possible which-path information this configuration results in a high degree of entanglement of the emitted photon pairs \cite{Jahnke2012,heinze17}, at least at low temperature \cite{Seidelmann2019}.

\section{Driven four-Level emitter}
\label{sec:FLE}

\subsection{Bare state picture}
\label{subsec:bare_state_pic}

\begin{figure}
\centering
\includegraphics[width=\columnwidth]{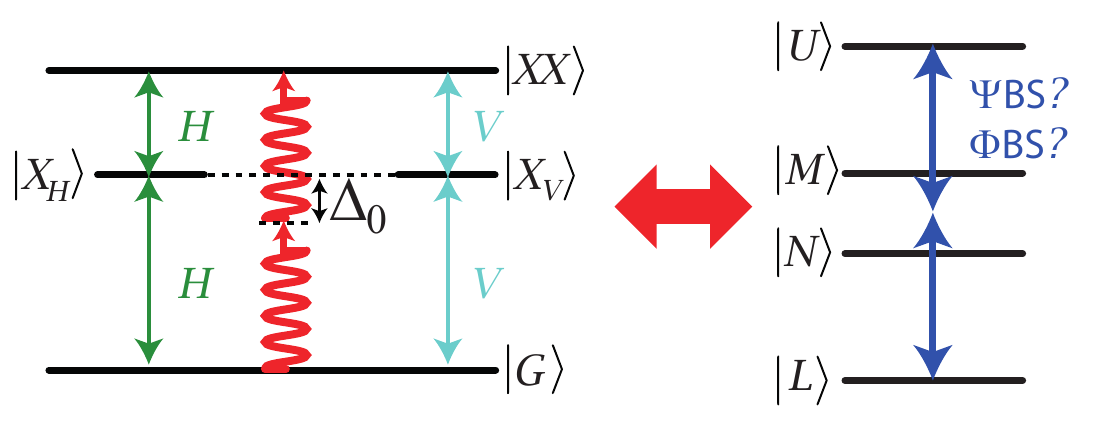}
\caption{Left: Sketch of the FLE including optical selection rules for transitions with either horizontally ($H$) or vertically ($V$) polarized light. In addition, an external laser field excites the system. Right: Sketch of the laser-dressed states.}
\label{fig:model}
\end{figure}

We consider an externally driven FLE embedded inside a microcavity, adopting the model from Ref.~\cite{munoz15}. The FLE comprises the energetic ground state $\KET{G}$ at energy $0$, two degenerate intermediate states $\KET{X_\text{H/V}}$ with energy $\hbar\omega_\text{X}$, and the upper state $\KET{XX}$ at energy $2\hbar\omega_\text{X}-E_\text{B}$. Note that it is quite common to find the state $\KET{XX}$ not exactly at twice the energy of the single excited states, which in quantum dots is known as the biexciton binding energy \cite{Orieux_entangled,Mermillod:16,Ota_2pht_emis_cavity_2011}. Optical transitions which involve the state $\KET{X_\text{H}}$ ($\KET{X_\text{V}}$) are evoked by horizontally (vertically) polarized light. Following Ref.~\cite{munoz15}, we assume the fine-structure splitting between these two intermediate states to be zero. A sketch of the FLE is shown in Fig.~\ref{fig:model}(left). The Hamiltonian of the FLE reads
\begin{eqnarray}
\label{eq:H_FLE}
\hat{H}_\text{FLE} &=& \hbar\omega_\text{X}\left( \OP{X_\text{H}}{X_\text{H}} + \OP{X_\text{V}}{X_\text{V}}\right)\\
	 &&+\left( 2\hbar\omega_\text{X} - E_\text{B} \right) \OP{XX}{XX} \notag .
\end{eqnarray}
The FLE is continuously driven by an external laser with frequency $\omega_\text{L}$ and driving strength $\Omega$. The laser driving is assumed to be linearly polarized, such that the $H$ and $V$ polarized transitions are driven with equal strength ensuring that there is no preferred polarization and, consequently, no which-path information is introduced by the external laser. In the frame co-rotating with the laser frequency $\omega_\text{L}$ the corresponding Hamiltonian reads
\begin{equation}
\label{eq:H_L}
\hat{H}_\text{L} = \Omega \left( \hat{\sigma}_\text{D} + \hat{\sigma}_\text{D}^\dagger \right); 
\qquad
\hat{\sigma}_\text{D} = \left( \hat{\sigma}_\text{H} + \hat{\sigma}_\text{V} \right) / \sqrt{2}
\end{equation}
with the transition operators 
\begin{subequations}
\begin{eqnarray}
\hat{\sigma}_\text{H} &=& \OP{G}{X_\text{H}} + \OP{X_\text{H}}{XX} , \\
\hat{\sigma}_\text{V} &=& \OP{G}{X_\text{V}} + \OP{X_\text{V}}{XX}\,.
\end{eqnarray}
\end{subequations}
We fix the laser frequency to $\hbar\omega_\text{L} = (2\hbar\omega_\text{X} - E_\text{B}) / 2$, such that the energetic detuning between emitter transitions and laser is set to 
\begin{equation}
	\Delta_0 := \hbar \left( \omega_\text{X} - \omega_\text{L} \right)= \frac{E_\text{B}}{2} .
\end{equation}
By this, we resonantly drive the two-photon transition between ground state $\KET{G}$ and upper state $\KET{XX}$. 

The FLE is embedded inside a microcavity and coupled to two orthogonal linearly polarized cavity modes with energies $\hbar\omega_\text{H}^c$ and $\hbar\omega_\text{V}^c$, which we assume to be energetically degenerate, i.e., $\omega_c := \omega_\text{H}^c = \omega_\text{V}^c$. The cavity mode is best defined with respect to the driving laser frequency (or the two-photon resonance to $\KET{XX}$) via the cavity laser detuning
\begin{equation}
	\Delta := \hbar \left( \omega_c - \omega_\text{L} \right) = \hbar\omega_c -\left(\hbar \omega_\text{X}- \Delta_0 \right) .
\end{equation}
The Hamiltonian describing the cavity modes and their interaction with the FLE reads
\begin{equation}
\hat{H}_\text{c} = \sum\limits_{\ell=H,V} \Delta \hat{a}_\ell^\dagger\hat{a}_\ell + \hat{H}_\text{FLE-c}.
\end{equation}
In matrix form, using the basis $\KET{XX}$, $\KET{X_\text{H}}$, $\KET{X_\text{V}}$, and $\KET{G}$, the interaction Hamiltonian is given as
\begin{equation}
\label{eq:H_C}
\hat{H}_\text{FLE-c} =
\begin{pmatrix}
0 &g\aH  & g\aV & 0 \\
g\aHd & 0 & 0 & g\aH \\
g\aVd & 0 & 0 & g\aV \\
0 & g\aHd & g\aVd & 0
\end{pmatrix} ,
\end{equation}
where the emitter-cavity coupling constant $g$ is assumed equal for all transitions. The bosonic operators $\hat{a}_\text{H/V}^\dagger$ ($\hat{a}_\text{H/V}$) create (annihilate) one cavity photon with frequency $\omega_c$ and $H/V$ polarization.
Note that $\hat{H}_\text{c}$ is again written in the rotating frame.
From the interaction Hamiltonian we can already see that in the un-driven situation the cascade from the state $\KET{XX}$ into the state $\KET{G}$ can only go via the emission of two $H$ or two $V$ polarized photons and therefore can result exclusively in the generation of $\Phi$BS entanglement. 

\subsection{Laser-dressed states}
\label{subsec:dressed_state_pic}

\begin{figure}
\centering
\includegraphics[width=\columnwidth]{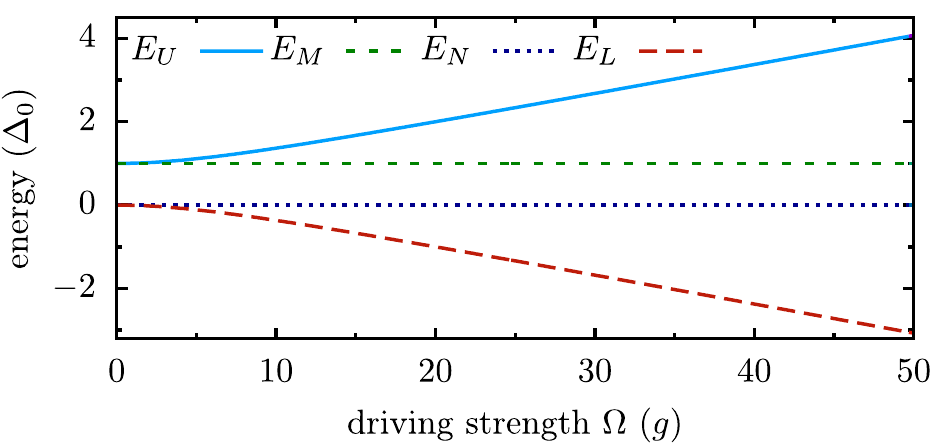}
\caption{Energies of the laser dressed states (in the units of the emitter-laser detuning $\Delta_0$) as a function of the driving strength $\Omega$ (in the units of the emitter-cavity coupling strength $g$).}
\label{fig:dressed}
\end{figure}

The creation of entangled two-photon states is facilitated by resonant transitions between quantum states of the FLE with the emission of two photons. Further analysis of the system dynamics reveals that such transitions take place not between the original FLE basis states but between the dressed states of the laser driven FLE, obtained by diagonalizing $\hat{H}_\text{FLE}+\hat{H}_\text{L}$. For the diagonalization we go into a frame rotating with the laser frequency $\omega_\text{L}$. The
eigenenergies of the dressed states read
\begin{subequations}
\begin{eqnarray}
E_\text{U} &=&  \frac{1}{2} \left( \Delta_0 +\sqrt{\Delta_0^2+ 8\Omega^2} \right)\\
E_\text{M} &=& \Delta_0 \\
E_\text{N} &=& 0 \\
E_\text{L} &=&  \frac{1}{2} \left( \Delta_0 - \sqrt{\Delta_0^2+ 8\Omega^2} \right)
\end{eqnarray}
\end{subequations}
and the corresponding laser-dressed states are
\begin{subequations}
\label{eq:def_dressed_states}
\begin{eqnarray}
\label{eq:def_U/L_state}
\KET{{U}} &=& c \left( \KET{G}+\KET{XX} \right) 
+ \tilde{c}\left( \KET{X_\text{H}}+\KET{X_\text{V}} \right) \\
\label{eq:def_M_state}
\KET{M} &=& \frac{1}{\sqrt{2}} \left( \KET{X_\text{H}}-\KET{X_\text{V}} \right)\\
\label{eq:def_N_state}
\KET{N} &=& \frac{1}{\sqrt{2}} \left( \KET{G}-\KET{XX} \right)\\
\KET{{L}} &=& \tilde{c} \left( \KET{G}+\KET{XX} \right) 
- c \left( \KET{X_\text{H}}+\KET{X_\text{V}} \right)
\end{eqnarray}
\end{subequations}
with the coefficients
\begin{eqnarray*}
 c &=&\dfrac{2\Omega}{\sqrt{8\Omega^2+ \left( \Delta_0 + \sqrt{\Delta_0^2+8\Omega^2} \right)^2}}\, , \quad 
\tilde{c} = \sqrt{\frac{1}{2}-c^2}\,.
\end{eqnarray*}
A sketch of the four laser-dressed states is given in Fig.~\ref{fig:model} (right panel). The dependence of the dressed state energies on the driving strength $\Omega$ is illustrated in Fig. \ref{fig:dressed}. The uppermost $\KET{U}$ and the lowest $\KET{L}$ states have contributions of all four original (bare) FLE states.  In the limiting case of strong driving the contribution coefficients $c$ and $\tilde{c}$ approach $1/2$.  On the other hand, the composition and energies of the intermediate dressed states $\KET{M}$ (``middle'') and $\KET{N}$ (``null'') are independent of $\Omega$.
In general, the laser-dressed states and the transition energies between them are functions of $\Omega$. Therefore, also the cavity frequency associated with a two-photon resonance between two given dressed states depends on the driving strength, the only exception being the resonance between the states $\KET{M}$ and $\KET{N}$.

The Hamiltonian describing the coupling to the cavity also changes profoundly by using the dressed state basis and now reads in the basis $\KET{U}$, $\KET{M}$, $\KET{N}$, $\KET{L}$
\begin{equation}
\label{eq:H_DS}
\begin{split}
&\hat{H}_{\text{DS-c}} =  \\
& g
\begin{pmatrix}
2\sqrt{2} c\tilde{c} \,\aDd & c \,\aAd & -\tilde{c} \,\aDd & \sqrt{2} (\tilde{c}^2-c^2) \aDd \\
c \,\aAd & 0 & \frac{-1}{\sqrt{2}} \aAd & \tilde{c} \,\aAd \\
\tilde{c} \,\aDd & \frac{1}{\sqrt{2}} \,\aAd & 0 &- c\,\aDd \\
\sqrt{2} (\tilde{c}^2-c^2)\aDd & \tilde{c} \,\aAd & c \,\aDd & -2\sqrt{2} c\,\tilde{c} \,\aDd
\end{pmatrix} + c.c.
\end{split}
\end{equation}
with $\aDd = (\aHd+\aVd)/\sqrt{2}$ and $\aAd = (\aHd-\aVd)/\sqrt{2}$ being the creation operators in the diagonal and anti-diagonal polarization, respectively. 

One notes that the two-photon transitions between the dressed states can follow different  pathways that connect those states.  Considering as an example the transition from $\KET{U}$ to $\KET{L}$, one path is to emit two photons with anti-diagonal polarization $A$ via the intermediate state $\KET{M}$, while another path is a self interaction within $\KET{U}$ and then a direct transition to $\KET{L}$ via emission of two diagonally $D$-polarized photons. This already indicates that due to the constant optical driving it is not clear a priori, which entanglement type occurs. We will show below that new types of entanglement become possible and analyze their respective strength. 

\subsection{Cavity losses and radiative decay}
\label{subsec:losses}
To account for cavity losses and radiative decay, present in every FLE-cavity system, we introduce Lindblad-type operators
\begin{equation}
\label{eq:Lindblad}
\LB{\hat{O}}{\Gamma}\,\hat{\rho} = \frac{\Gamma}{2} \left( 2\hat{O}\hat{\rho}\hat{O}^\dagger - \hat{\rho}\hat{O}^\dagger\hat{O} - \hat{O}^\dagger\hat{O}\hat{\rho} \right),
\end{equation}
where $\hat{O}$ is the system operator associated with a loss process with corresponding loss rate $\Gamma$ in the bare state system. The dynamics of the statistical operator of the system $\hat{\rho}$ is then determined by the Liouville-von Neumann  equation
\begin{equation}
\label{eq:Eq_of_Motion}
\begin{split}
\ddt\hat{\rho} =& \mathcal{L}\hat{\rho} := -\frac{i}{\hbar} \left[ \hat{H},\hat{\rho} \right] \\
&+ \sum\limits_{\ell=H,V} \Big\lbrace \LB{\hat{a}_\ell}{\kappa} + \LB{\OP{G}{X_\ell}}{\gamma} + \LB{\OP{X_\ell}{XX}}{\gamma} \Big\rbrace \hat{\rho},
\end{split}
\end{equation}
where $[\cdot,\cdot]$ denotes the commutator, $\kappa$ is the cavity loss rate, and $\gamma$ the radiative decay rate. The complete system Hamiltonian $\hat{H}$ includes all contributions discussed in Sec.~\ref{subsec:bare_state_pic}. The system is assumed initially in the ground state $\KET{G}$ without any cavity photons. Note that we performed all numerical calculations in the rotating frame with the laser frequency $\omega_\text{L}$ and use the bare state system, while for the interpretation the dressed state picture is advantageous. 

The parameter values used in our simulations are listed in Table~\ref{tab:Fixed_Parameters}, where we followed Ref.~\cite{munoz15}. The frequency of the cavity mode is taken to $\hbar \omega_c = 1.5$ eV. The adopted parameter values correspond to a high quality cavity resonator with $Q = 1.5 \times 10^5$.

\begin{table}
\centering
\caption{Fixed system parameters used in the calculations.}
\label{tab:Fixed_Parameters}
\begin{tabular}{l c c}
\hline\hline
Parameter & & Value\\
\hline
Emitter-cavity coupling strength & $g$ & 0.051 meV \\
Detuning & $\Delta_0$ & $20g=1.02$~meV \\
Cavity loss rate & $\kappa$ & $0.1g/\hbar \approx 7.8$~$\mathrm{ns^{-1}}$\\
Radiative decay rate & $\gamma$ & $0.01g/\hbar \approx 0.78$~$\mathrm{ns^{-1}}$\\
\hline\hline
\end{tabular}
\end{table}

\section{Photon entanglement}
\label{sec:classification_entanglement}

\subsection{Two-photon density matrix}
\label{subsec:2PhtDenMat}
The basis for quantifying the degree of entanglement is the determination of the two-photon density matrix $\rho^\text{2p}$. Experimentally, $\rho^\text{2p}$ can be reconstructed using methods of quantum state tomography \cite{QuantumStateTomography}, a technique based on polarization-resolved two-time coincidence measurements. The detected signals are proportional to the two-time correlation functions
\begin{equation}
\label{eq:G2}
G_{jk,lm}^{(2)}(t,\tau) = \left\langle \hat{a}_j^\dagger(t)\hat{a}_k^\dagger(t+\tau)\hat{a}_m(t+\tau)\hat{a}_l(t) \right\rangle,
\end{equation}
where $\lbrace j,k,l,m\rbrace\in\lbrace H,V\rbrace$, $t$ is the real time when the first photon is detected, and $\tau$ the delay time between the detection of the first and the second photon. Note that in experiments one typically measures photons that have already left the cavity. However, considering the out-coupling of light out of the cavity to be a Markovian process, Eq.~\eqref{eq:G2} can also describe $G_{jk,lm}^{(2)}(t,\tau)$ measured outside the cavity \cite{Kuhn:16,Different-Concurrences:18}.

In experiments data is typically averaged over finite real time and delay time windows. Thus, the experimentally reconstructed two-photon density matrix is calculated as \cite{munoz15,Different-Concurrences:18}
\begin{equation}
\label{eq:rho_jklm_Def}
\rho_{jk,lm}^\text{2p}(\tau) = \frac{\overline{G}_{jk,lm}^{(2)}(\tau)}{\mathrm{Tr}\left\lbrace \overline{G}^{(2)}(\tau) \right\rbrace},
\end{equation}
where $\overline{G}^{(2)}$ is the time-averaged correlation with
\begin{equation}
\label{eq:G_jklm}
\overline{G}_{jk,lm}^{(2)}(\tau) = \frac{1}{\Delta t\,\tau} \int\limits_{t_0}^{t_0+\Delta t}\mathrm{d}t \int\limits_{0}^{\tau}\mathrm{d}\tau^\prime G_{jk,lm}^{(2)}(t,\tau^\prime)\,.
\end{equation}
Here, $\tau$ ($\Delta t$) is the delay time (real time) window used in the coincidence measurement and $t_0$ is its starting time. The trace $\mathrm{Tr}\lbrace\cdot\rbrace$ is introduced for normalization. For simplicity we refer to $\rho^\text{2p}$ as the two-photon density matrix in the following.

Throughout this work we calculate the two-photon density matrix for the system that reached its steady state so that the $t$-average is independent of $t_0$ and $\Delta t$. The steady state of the system $\hat{\rho}_s$ is defined by $\ddt\hat{\rho}_s = \mathcal{L}\hat{\rho}_s = 0$. This state is obtained numerically by letting the system evolve in time until its density matrix becomes stationary. We will further  set $\tau=50$~ps, which is a realistic value for the delay time window used in experiment \cite{StevensonPRL2008}. More details on the calculation of the two-time correlation functions for systems including Markovian loss processes can be found in Ref.~\onlinecite{multi-time}.

\subsection{Concurrence}
\label{subsec:concurrence}

Using the two-photon density matrix  we determine the corresponding concurrence $C$ \cite{Wootters1998}, which is a widely accepted measure for the degree of entanglement of a bipartite system. The concurrence is calculated from a given two-photon density matrix $\rho^\text{2p}$ according to \cite{EdV,Wootters1998,QuantumStateTomography}
\begin{equation}
\label{eq:Con_definition}
C = \max\left\lbrace 0, \sqrt{\lambda_1} - \sqrt{\lambda_2} - \sqrt{\lambda_3} - \sqrt{\lambda_4} \right\rbrace
\end{equation}
where $\lambda_j$ are the (real and positive) eigen-values in decreasing order, $\lambda_1 \geq \lambda_2 \geq \lambda_3 \geq \lambda_4$, of the matrix
\begin{equation}
\label{eq:def_M_matrix}
M = \rho^\text{2p} \, T\, (\rho^{\text{2p}})^\ast \, T ,
\end{equation}
where $T$ is an anti-diagonal matrix of rank 4 with elements $\left\lbrace -1,1,1,-1\right\rbrace$ and $(\rho^\text{2p})^{\ast}$ is the complex conjugated two-photon density matrix.
In the standard situation without driving, where only a $\Phi$BS $\KET{\Phi_\pm}$ can be generated, the full expression for the concurrence reduces to $C = 2\ABS{\rho^\text{2p}_{HH,VV}}$. Thus, the degree of entanglement is closely related to the corresponding coherences in the two-photon density matrix.
Note that like the two-photon density matrix $\rho^\text{2p}(\tau)$ also the concurrence $C(\tau)$ depends on the measurement window $\tau$.
A finite delay time window $\tau$ is necessary for the detection of $\Psi$BS entanglement since the two contributions that build up $\KET{\Psi_+}$ in Eq.~\eqref{eq:def_bell_ent_state} can only be distinguished if the two photons are detected at different times \cite{munoz15}.

For the numerical calculation of the concurrence we use the following procedure: First, following Ref.~\cite{multi-time}, the averaged two-time photon correlation $\overline{G}^{(2)}$ is calculated. This quantity is then used to obtain the time-averaged two-photon density matrix in Eqs.~\eqref{eq:rho_jklm_Def}.
Finally from the two-photon density matrix the concurrence is determined according to Eq.~\eqref{eq:Con_definition}. Note that we do not use any further approximations in the calculation of $\overline{G}^{(2)}$.

\section{Two-photon transition between upper and lower dressed state}
\label{sec:transition_2p_UL}

\begin{figure*}[t!]
\centering
\includegraphics[width=\textwidth]{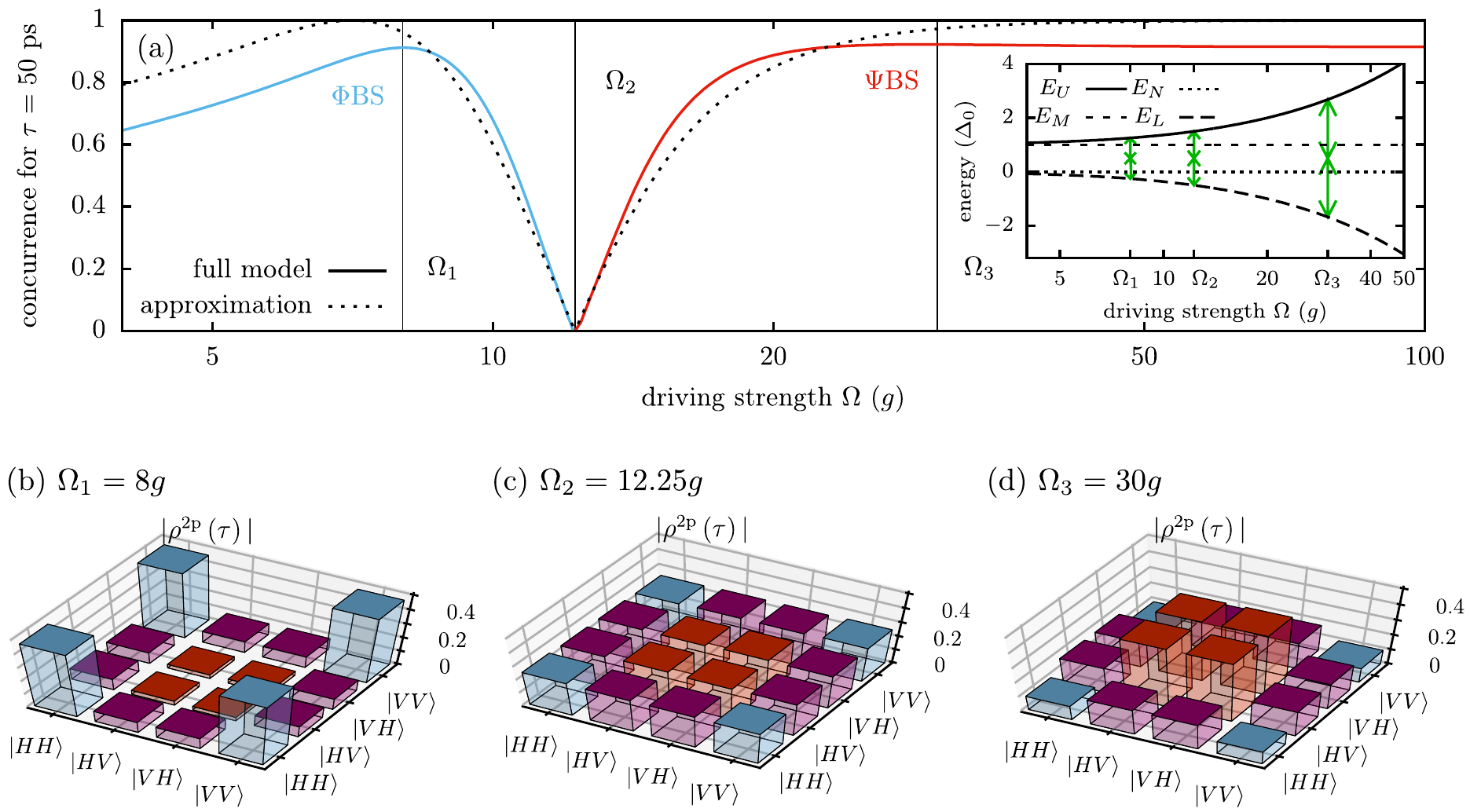}
\caption{(a) Concurrence as function of the laser driving strength $\Omega$ for the full model (solid line) and the analytic approximation $C(r)$ presented in Eq.~\eqref{eq:C_von_r} (dotted line). Inset: Dressed state energies as a function of the driving strength and the two-photon resonant cavity modes (green arrows) for three selected $\Omega$ values. (b)-(d) Absolute value of the two-photon density matrix $\vert\rho^\text{2p}(\tau)\vert$ for driving strength (a) $\Omega_1=8g$, (b) $\Omega_2=12.25g$ and (c) $\Omega_3=30g$ (indicated by vertical lines in (a)).}
\label{fig:transition}
\end{figure*}

The emission of entangled two-photon states is associated with two photon transitions between the dressed FLE states.
The dressed FLE states feature two-photon emissions, which are largest every time the cavity frequency is tuned in resonance with a possible two-photon transition, i.e., when twice the photon energy (here $\Delta$) is equal to the transition energy between the dressed state pairs.
Therefore, the analysis is focused on these resonant situations.

We start our analysis with the case where the cavity photons are in resonance with the transition between the states $\KET{U}$ and $\KET{L}$, i.e., the cavity frequency is always tuned such that
\begin{equation}
\label{eq:delta_UL}
	\Delta=\frac{E_\text{U} - E_\text{L}}{2}  = \frac{1}{2}\sqrt{\Delta_0^2 + 8\Omega^2} .
\end{equation}
Notice, that keeping this condition requires the cavity frequency $\omega_c$ to change with the driving strength  $\Omega$. This resonance for a driven FLE was considered in earlier works \cite{munoz15}, where a possibility to achieve a high degree of $\Psi$BS entanglement was demonstrated. Here we demonstrate that $\Psi$BS entanglement is not the only type of two-photon entanglement that can be obtained. It will be shown that by varying the driving strength (while keeping the system at the considered resonance) the FLE can reach the domain of $\Phi$BS entanglement, separated from that of the $\Psi$BS by a special critical point of zero concurrence.

\subsection{Transition between $\mathbf{\Phi}$BS and $\mathbf{\Psi}$BS entanglement}
\label{subsec:transition_del_Valle}

The concurrence as a function of the driving strength $\Omega$ is shown in Fig.~\ref{fig:transition}(a), where the inset illustrates the  resonance in question. In full agreement with earlier calculations \cite{munoz15} one observes $\Psi$BS entanglement when the driving is strong. However, when the driving strength is lowered the entanglement changes its type to $\Phi$BS entanglement. A sharp transition between the two types occurs at a special critical point $\Omega\approx12.25g$ where the concurrence is exactly zero.
The $\Phi$BS entanglement obtained for weak driving reflects the fact that for small $\Omega$ the system approaches the undriven case. Recalling that $\Psi$BS entanglement has been found in Ref.~\cite{munoz15} for higher $\Omega$, it is clear that a transition has to take place in between.

More insight into the entanglement change is obtained by calculating the corresponding two-photon density matrices as presented in Fig.~\ref{fig:transition}(b) for the driving strength $\Omega_1 = 8g$ and Fig.~\ref{fig:transition}(d) $\Omega_3 = 30g$. At $\Omega_1$ the occupations of the states $\KET{HH}$ and $\KET{VV}$ and their coherence clearly dominate over the remaining elements representing $\Phi$BS entanglement. A very different behavior is found at $\Omega_3=30g$, where the occupations of the states $\KET{HV}$ and $\KET{VH}$ and the corresponding coherences exhibit the highest values and, consequently, are associated with $\Psi$BS entanglement.

Let us now focus on the special point at $\Omega_2=12.25g$. The two-photon density matrix at the special point, shown in Fig.~\ref{fig:transition}(c), reveals that the concurrence does not vanish because of the absence of coherences. On the contrary, all coherences are close to their maximal possible value of about $0.25$. Further analysis reveals that the corresponding two-photon state is
\begin{equation}
\label{eq:psi_special_point}
\begin{split}
\KET{\psi_{\text{sp}}} =& \frac{1}{2} \left( \KET{HH} - \KET{HV} - \KET{VH} + \KET{VV} \right) \\
 =& \frac{1}{\sqrt{2}} \left( \KET{H_1} - \KET{V_1} \right) \frac{1}{\sqrt{2}} \left( \KET{H_2} - \KET{V_2} \right).
\end{split}
\end{equation} 
Remarkably, this is a pure state and $\KET{\psi_{\text{sp}}}$ can be factorized into a product of two one-photon states describing the first and second detected photon, respectively (indicated by 1 and 2). Since $\KET{\psi_{\text{sp}}}$ can be factorized, it is not entangled and, thus, the concurrence vanishes at this point. Therefore, instead of a direct transition from high $\Phi$BS to high $\Psi$BS entanglement the system passes through this special point with vanishing degree of entanglement.

We note that the special point occurs at a distinct resonance condition. Beside the two-photon transition between the two outermost dressed states, also the one-photon process between the intermediate states $\KET{M}$ and $\KET{N}$ state becomes resonant.

\subsection{Effective Hamiltonian of the system at the resonance}
\label{subsec:SWT}

In order to understand the underlying physics of the crossover between the entanglement types we derive an effective Hamiltonian that describes the most relevant transition processes involving the $\KET{U}$ and $\KET{L}$ states. To be more specific, we account only for the uppermost state without photons $\KET{U,0,0}$ and the lowest states with two photons $\KET{L,1,1}$, $\KET{L,2,0}$, and $\KET{L,0,2}$. Here, $\KET{\chi,n_\text{H},n_\text{V}}$ is the product state of $\KET{\chi}\in\lbrace\KET{U},\KET{M},\KET{N},\KET{L}\rbrace$ and the photon number state for $H$ and $V$ polarization.

\begin{figure}[t]
\centering
\includegraphics[width=\columnwidth]{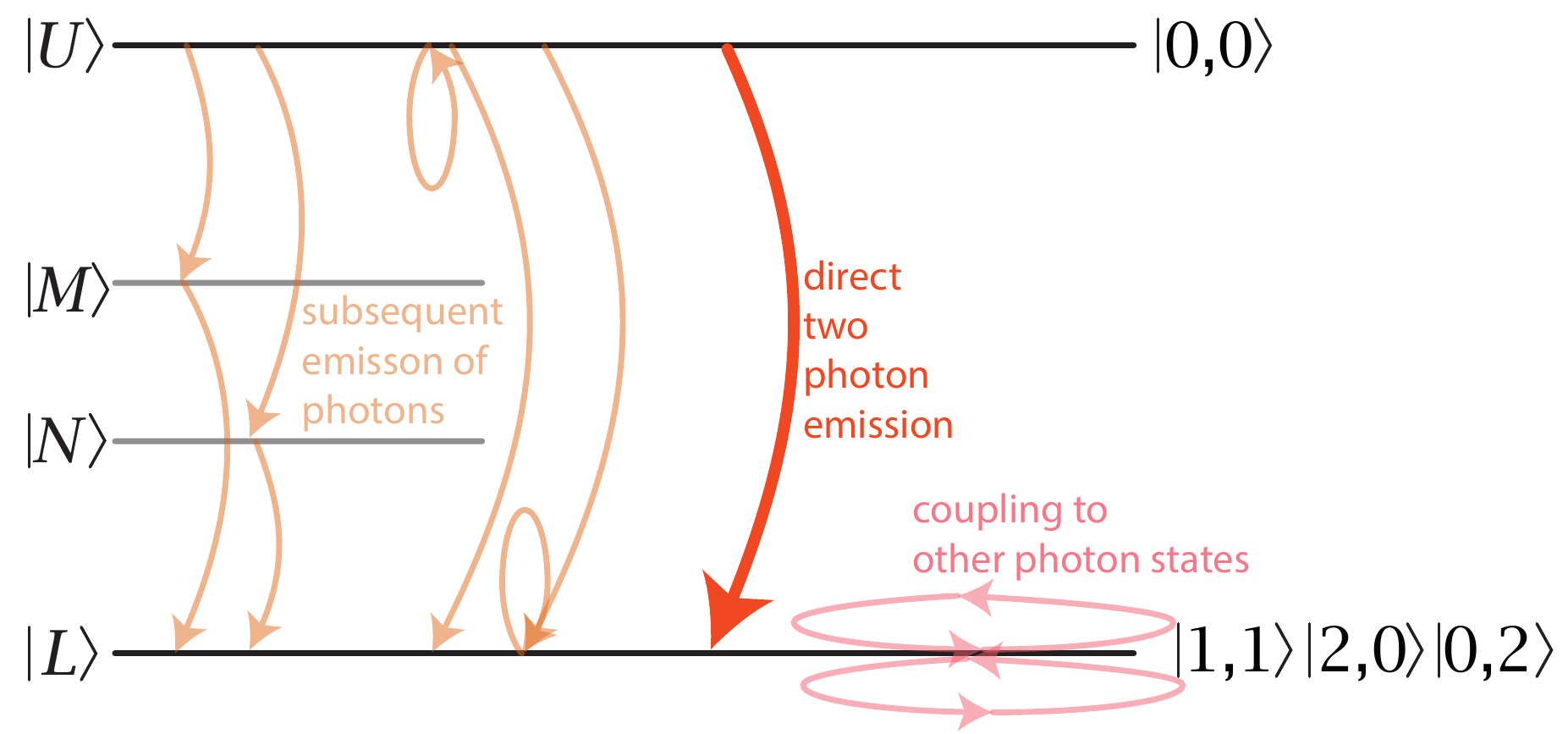}
\caption{Schematic depiction of the possible transitions connecting $\KET{U,0,0}$ to the two-photon states $\KET{L,1,1}$, $\KET{L,2,0}$, and $\KET{L,0,2}$. All but the direct two-photon emission process (bold orange arrow) are eliminated in the Schrieffer-Wolff transformation.}
\label{fig:SchriefferWolffTrafo}
\end{figure}

Besides the direct two-photon transitions, there are several other possibilities to go from the initial to the final states. One example are subsequent one photon transitions, either going via one of the intermediate states or by a self-interaction and then a one-photon process. Also, from the final states, a sequential photon emission and absorption (or the other way around) can take place. These processes are depicted in Fig.~\ref{fig:SchriefferWolffTrafo}. Therefore, the states mentioned above are coupled to a bunch of other states, namely the one-photon states $\KET{\chi,1,0}$, $\KET{\chi,0,1}$ and the three-photon states $\KET{\chi,3,0}$, $\KET{\chi,2,1}$, $\KET{\chi,1,2}$, and $\KET{\chi,0,3}$ (with $\chi\in\lbrace U,M,N,L\rbrace$), while the latter can be reached in sequential emission/absorption processes.

Using a Schrieffer-Wolff transformation, it is now possible to encode these transitions into a single matrix, acting only within the basis spanned by the direct two-photon transitions, i.e., $\KET{U,0,0}$, $\KET{L,1,1}$, $\KET{L,2,0}$, and $\KET{L,0,2}$  \cite{Winkler,LossDivincenzo_SW}. A Schrieffer-Wolff transformation thereby performs a block-diagonalization, which decouples the desired states from the rest. This is reasonable, because the removed states are strongly off-resonant in this situation and, thus, represent a small perturbation. More details on the Schrieffer-Wolff transformation can be found in App.~\ref{app:SWT}.

After the Schrieffer Wolff-transformation, which is treated within the photon number states, we afterwards perform additionally a basis transformation to rotate the system partially into the Bell basis with $\left\lbrace\KET{U,0,0}, \KET{L,1,1},\KET{L,\Phi_+},\KET{L,\Phi_-}\right\rbrace$. In this representation $\KET{L,1,1}$ corresponds to the possibility of $\Psi$BS entanglement, where two photons are generated such that one is $H$- and the other $V$-polarized. However, without further analysis, we cannot distinguish between $\Psi_{\pm}$BS entanglement. The effective Schrieffer-Wolff Hamiltonian is then given by
\begin{eqnarray}
\label{eq:result_2p_UL_v3}
&&\hat{\tilde{H}}_{\text{UL}}^{(2)} 
= g ^2  \begin{pmatrix}
\delta^{\text{UL}} & \gamma_1^{\text{UL}} & -\gamma_2^{\text{UL}} & 0\\
\gamma_1^{\text{UL}} & -\delta^{\text{UL}} & \alpha^{\text{UL}} & 0\\
-\gamma_2^{\text{UL}} & \alpha^{\text{UL}} & -\delta^{\text{UL}} & 0\\
0 & 0 & 0 & -\delta^{\text{UL}}
\end{pmatrix}
\end{eqnarray}
with
\begin{eqnarray}
\delta^{\text{UL}} &=& \left(\tilde{c}^2-c^2\right) \left(\frac{2}{\Delta_0}+\frac{4}{\dcL{UL}}\right) \notag \\
\gamma_1^{\text{UL}} &=& 4c\tilde{c}\frac{1}{\Delta_0}  - 16 c\tilde{c}\left(\tilde{c}^2- c^2\right)\frac{1}{\dcL{UL}} \notag \\ 
\gamma_2^{\text{UL}} &=& 16 c\tilde{c}\left(\tilde{c}^2- c^2\right)\frac{1}{\dcL{UL}} \notag \\
\alpha^{\text{UL}} &=& \frac{1}{\Delta_0} - \left(1 -16 c^2 \tilde{c}^2\right)\frac{1}{\dcL{UL}}  \,. \notag 
\end{eqnarray}
The given expressions contain only the most important contributions. The full expressions can be found in App.~\ref{app:subsec:2p_UL}.
It is interesting to note that the coefficients $\gamma_{1/2}^{\text{UL}}$ stem from the subsequent emission of two single photons (faded orange arrows in Fig.~\ref{fig:SchriefferWolffTrafo}) and simultaneous two-photon emission, while $\alpha^{\text{UL}}$ accounts for the fact that from the two photon states, coupling to higher (lower) photon states can take place and therefore couple different types of two-photon states  (faded red arrows in Fig.~\ref{fig:SchriefferWolffTrafo}). An example for the latter case is the coupling of $\vert L, 2,0\rangle \to \vert L,2,1\rangle$, followed by a photon number reduction via $\vert L,2,1\rangle \to \vert L,1,1\rangle$ illustrating why different two-photon states are coupled. 

\begin{figure}[t]
\centering
\includegraphics[width=0.9\columnwidth]{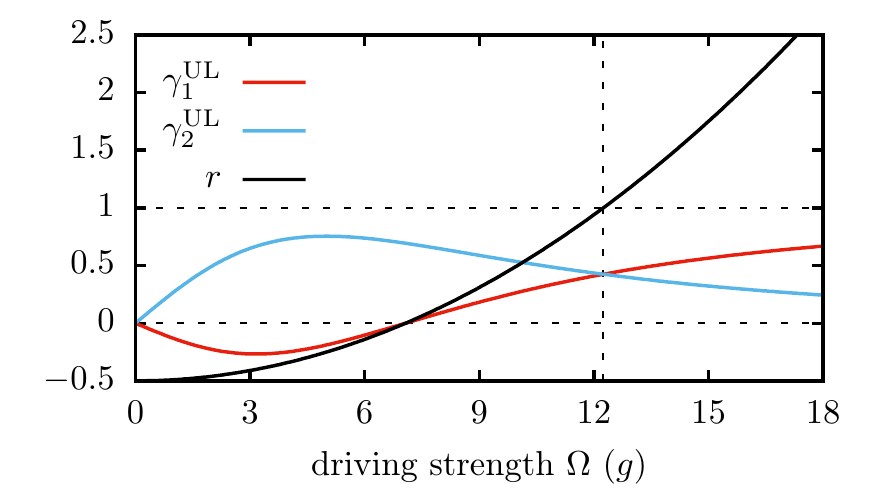}
\caption{Effective coupling constantes $\gamma_1^\text{UL}$ and $\gamma_2^\text{UL}$ and the ration $r = \gamma_1^\text{UL}/\gamma_2^\text{UL}$ as function of driving strength $\Omega$. }
\label{fig:coeff}
\end{figure}

From this Hamiltonian, we can now deduce which type of entanglement is created: First of all we find that the state $\KET{L,\Phi_-}$ is decoupled, such that we see that photons with this type of entanglement are not created. In contrast, the initial state $\KET{U,0,0}$ is coupled to the $\KET{L,\Phi_+}$ state via $\gamma_2^{\text{UL}}$ and to the state $\KET{L,1,1}$ via $\gamma_1^{\text{UL}}$. Therefore in principle both $\Phi$BS and $\Psi$BS entanglement can be created. The different types of entangled states are coupled via the coefficient $\alpha^{\text{UL}}$, however, we will for now neglect this coupling (see discussion at the end of the next section). Which type of entanglement dominates depends on the ratio
\begin{equation}
	r=\frac{\gamma_1^{\text{UL}}}{\gamma_2^{\text{UL}}} = 4\left(\frac{\Omega}{\Delta_0}\right)^2-\frac{1}{2} .
\end{equation}
This means, we obtain preferably $\Phi$BS entanglement, when $\gamma_2^{\text{UL}}>\gamma_1^{\text{UL}}$ (or $\ABS{r}<1$), and preferably $\Psi$BS entanglement if $\gamma_2^{\text{UL}}<\gamma_1^{\text{UL}}$ (or $\ABS{r}>1$). Figure~\ref{fig:coeff} displays the ratio $r$ as well as the couplings $\gamma_1^\text{UL}$ and $\gamma_2^\text{UL}$ as a function of the driving strength $\Omega$. Indeed, $r=1$ corresponds to $\Omega_{\text{sp}} = \sqrt{3/8}\Delta_0$ and we obtain our special point, when both types of entanglement are occurring with equal weight and we have zero concurrence since their superposition results in a factorizable state.  \\

\subsection{Approximate two-photon density matrix}
\label{subsec:approximate_result}

Further insight is obtained by calculating the two-photon density matrix assuming the delay window $\tau$ is small and can be neglected so that
\begin{equation}
\label{eq:tpdm_small_tau}
\begin{split}
\rho^\text{2p}_{jk,lm}(\tau)  &\approx  \mathcal{N} \Tr{\hat{a}_m\, \hat{a}_l\, \hat{\rho}_{\text{s}}\,\hat{a}_j^\dagger\,\hat{a}_k^\dagger}
\end{split}
\end{equation}
where $\mathcal{N}$ is a normalization constant and $\hat{\rho}_{\text{s}}$ describes the steady state of the system. Note that only states with at least two photons inside the cavity contribute to the two-photon density matrix. Neglecting the coupling $\alpha^\text{UL}$ in the effective Hamiltonian~\eqref{eq:result_2p_UL_v3} and performing another basis transformation, one finds that the only two-photon state coupled to $\KET{U,0,0}$ is
\begin{equation}
\KET{\psi_\text{s}} = \frac{1}{\sqrt{{\left(\gamma_1^\text{UL}\right)}^2+{\left(\gamma_2^\text{UL}\right)}^2}} \left( {\gamma_1^\text{UL}}\KET{L,1,1}-{\gamma_2^\text{UL}}\KET{L,\Phi_+} \right).
\end{equation}
Therefore, in this approximation, also the contribution to the steady state which contains two photons inside the cavity should be proportional to $\KET{\psi_\text{s}}$. Consequently, the approximate normalized two-photon density matrix can be calculated by inserting $\rho_s = \KET{\psi_\text{s}}\BRA{\psi_\text{s}}$ into Eq.~\eqref{eq:tpdm_small_tau} which results in
\begin{equation}
\label{eq:rho_approx}
\rho^\text{2p}_\text{approx} = 
\frac{1}{2(1+r^2)}
\begin{pmatrix}
1 & -r & -r & 1\\
-r & r^2 & r^2 & -r \\
-r& r^2 & r^2 & -r \\
1 & -r & -r & 1
\end{pmatrix} ,
\end{equation}
For this simplified density matrix, we can analytically calculate the concurrence $C$ [Eq.~\eqref{eq:Con_definition}] to
\begin{equation}
\label{eq:C_von_r}
C(r) = \frac{\ABS{1-r^2}}{1+r^2} .
\end{equation}
In Fig.~\ref{fig:transition}(a) the approximate result $C(r)$ is included as a dotted line. The approximate solution agrees quite well with the numerical results. This underlines the idea that the concurrence depends essentially on the ratio $r$. Also for the approximate solution we have the special point at $r=1$ and the regions of high entanglement and the corresponding type of entanglement can be directly extracted from the analytical result.  Below the special point we have $\ABS{r}<1$, therefore, $r^2<|r|$, resulting in a density matrix of $\Phi$BS entanglement. The maximum concurrence value appears around $\Omega = \frac{1}{2\sqrt{2}}\Delta_0 \approx 7.1g$ where the ratio $r$ passes through zero. Above $\Omega_{\text{sp}}$, we have $r\geq 1$ and $r^2>r$. Thus, in this regime one obtains $\Psi$BS entanglement in the two-photon density matrix.

We now discuss the deviations between the numerical and the approximate result for the concurrence. One obvious reason for the difference is the obmission of the coupling between the two-photon states (via one- or three-photon states), as indicated by $\alpha^{\text{UL}}$ in Eq.~\eqref{eq:result_2p_UL_v3}. This coupling mixes $\Phi$BS and $\Psi$BS, such that in the full model, the total obtained concurrence is reduced. Nonetheless, neglecting $\alpha^{\text{UL}}$ for the analysis is reasonable, when taking the cavity losses into account. By analysing the values of $\alpha^{\text{UL}}$ and $\gamma^{\text{UL}}$, we find that these are always smaller than the cavity loss rate $\kappa$. This means that the losses relax the system before the coupling between the different photon states becomes efficient. Another reason for the deviations is that for low driving strength values, other transitions between the laser-dressed states besides the discussed direct two-photon one become important as they get closer to resonance.

\section{Entanglement at the other Two-photon transitions}
\label{sec:concurrence_vs_detuning}

\begin{figure*}
\centering
\includegraphics[width=\textwidth]{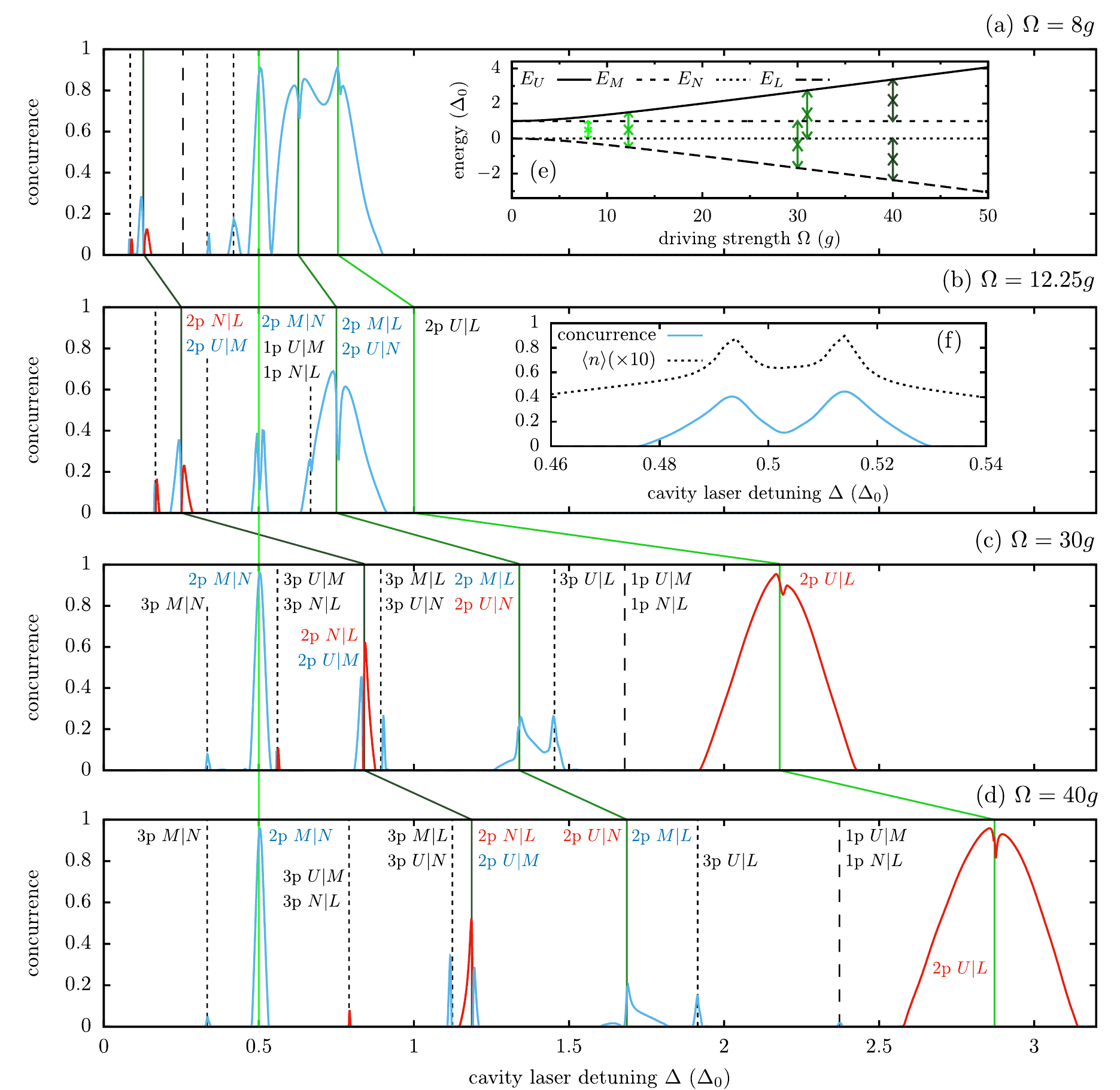}
\caption{Concurrence as function of the cavity laser detuning $\Delta$ for fixed values of the external laser driving (a) $\Omega = 8g$, (b) $\Omega = 12.25g$, (c) $\Omega = 30g$, and (d) $\Omega = 40g$. The color code indicates the type of entanglement: blue curves symbolizes $\Phi$BS and red curves are $\Psi$BS entanglement. The vertical lines mark the position of photon resonances labeled by  $n$p $\chi_1\vert\chi_2$. (e) Energy of the laser-dressed states as a function of the driving strength $\Omega$ marking the four selected two-photon resonance conditions which correspond to the two-photon resonances of the same color in panels (a)-(d). (f) Concurrence and mean photon number $\langle n\rangle$ for $\Omega=12.25g$ in the vicinity of $\Delta=\Delta_{\text{MN}}/2$. 
}
\label{fig:concurrence_vs_detuning}
\end{figure*}

Having discussed the transition between $\KET{U}$ and $\KET{L}$, we now want to examine the behavior of the other two-photon resonances. In particular, there are three other two-photon resonances matching the transitions between the corresponding dressed states (given by $\Delta_{\chi_1\chi_2}=E_{\chi_1}-E_{\chi_2}$) in the system at
\begin{eqnarray}
	\frac{\dcL{UM}}{2} = \frac{\dcL{NL}}{2}
	&=& \frac{1}{4}\left(\sqrt{\Delta_0^2+ 8\Omega^2}- \Delta_0 \right) \notag \\
	\frac{\dcL{UN}}{2} = \frac{\dcL{ML}}{2}
	&=& \frac{1}{4}\left(\sqrt{\Delta_0^2+ 8\Omega^2} + \Delta_0 \right) \notag \\
	\frac{\dcL{MN}}{2} &=&
	  \frac{\Delta_0}{2} \notag .
\end{eqnarray}
Therefore, to sweep through the respective resonances, we now fix the driving strength and vary the cavity laser detuning $\Delta$. The corresponding concurrence is calculated and the results are shown in Fig.~\ref{fig:concurrence_vs_detuning} for four different driving strength $\Omega=8g$, $12.25g$, $30g$, and $40g$.

The type of entanglement is encoded in the color: Blue lines are for $\Phi$BS and red lines for $\Psi$BS entanglement. On first sight, we find that both types of entanglement occur when we vary $\Delta$. In addition to a strong concurrence at the four two-photon resonances, we find several other cavity detuning values with non-vanishing concurrence. We can attribute these to the one-photon resonances $U|M$ and $N|L$ and several three-photon resonances, which occur between the respective states. Accordingly, we have labeled all resonances by $n$p $\chi_1\vert\chi_2$, which denotes the $n$-photon resonances between the laser-dressed states $\KET{\chi_1}$ and $\KET{\chi_2}$. 

Figure~\ref{fig:concurrence_vs_detuning}(e) shows the dressed states as a function of the driving strength and we used colored arrows to mark the different two-photon resonances. The same colors are used to indicate the position of the two-photon resonances in Figs.~\ref{fig:concurrence_vs_detuning}(a)-(d). Before we will go through the two-photon resonances one-by-one (note that we already discussed the 2p~$U\vert L$ resonance), let us briefly remark some general findings:

While some $n$-photon transitions are always associated with the same type of entanglement, others can change from one to the other. This change may happen as a result of changing the cavity laser detuning or the driving strength. Furthermore, in between some of the resonance conditions the concurrence value stays at a finite level, whereas it passes through zero in other situations. A striking feature is the appearance of a second special point with vanishing concurrence between regions of high entanglement when the cavity laser detuning is approximately $\Delta \approx \dcL{UM}/2 =\dcL{NL}/2$, which we will discuss in detail in Sec.~\ref{subsubsec:2p_UM}.

Next, we will go through the two-photon resonances one-by-one. For each two-photon resonance we perform a Schrieffer-Wolff transformation, followed by a rotation of the states, such that each Hamiltonian in the following is given in the basis
\begin{equation}
	\left\{ \KET{\chi_1,0,0}, \KET{\chi_2,1,1}, \KET{\chi_2,\Phi_+},\KET{\chi_2,\Phi-}\right\}
\end{equation}
with $\chi_1$ being the higher energy state and $\chi_2$ being the lower energy state of the 2p $\chi_1\vert\chi_2$ resonance. More details on the Schrieffer-Wolff transformation are given in App.~\ref{app:SWT}.

\subsection{Two-photon~$M\vert N$ resonance}
\label{subsubsec:2p_MN}
We start by looking at 2p~$M\vert N$, which is the only two-photon transition for which the resonance condition does not depend on the driving strength. The corresponding transitions are marked by a light green line in Fig.~\ref{fig:concurrence_vs_detuning}. At this resonance the concurrence always displays $\Phi$BS entanglement. While the concurrence is mostly maximal at the resonance,  we find a decrease in strength at the maximum at $\Omega=12.25g$.

We use the Schrieffer-Wolff transformation to obtain the effective Hamiltonian
\begin{eqnarray}
\label{eq:result_2p_MN}
&&\hat{\tilde{H}}_{\text{MN}}^{(2)} 
= g^2 
\begin{pmatrix}
\delta^{\text{MN}} & 0 & 0 & \gamma_2^{\text{MN}}\\
0 & -\delta^{\text{MN}} & -\delta^{\text{MN}} & 0\\
0 & -\delta^{\text{MN}} & -\delta^{\text{MN}} & 0 \\
\gamma_2^{\text{MN}} & 0 & 0 & -\delta^{\text{MN}}
\end{pmatrix}
\end{eqnarray}
with
\begin{eqnarray}
\delta^{\text{MN}} &=& 2\left(\tilde{c}^2-c^2\right)\frac{1}{\dcL{UL}} \notag \\
\gamma_2^{\text{MN}} &=& -4\,c\,\tilde{c}\frac{1}{\dcL{UL}}. \notag 
\end{eqnarray}
Note that these are shortened expressions and the full expressions can be found in App.~\ref{app:subsec:2p_MN}.
From the Hamiltonian, it is obvious that the initial state is only coupled to the final state $\KET{N,\Phi_-}$, while the other two-photon states become uncoupled. This is in agreement with Fig.~\ref{fig:concurrence_vs_detuning}, where we only find $\Phi$BS at the 2p~$M\vert N$ resonance. 

The smaller height in concurrence at $\Omega=12.25g$ (see also Fig.~\ref{fig:concurrence_vs_detuning}(f)), can be traced back to the occurrence of several resonance conditions at the same driving strength, in particular the one-photon transitions 1p~$U\vert M$ and 1p~$N\vert L$. This is confirmed by looking at the mean photon number $\langle n\rangle = \langle\aHd\aH+\aVd\aV\rangle$ as displayed in Fig.~\ref{fig:concurrence_vs_detuning}(f). The alignment of several resonance conditions causes the peak to split into two separate resonances, as indicated by the mean photon number. Due to the additional one-photon resonances three-photon states with all four possible combinations of polarized photons gain a noticeable population and the extracted (two-photon) coherence $\rho^\text{2p}_{\text{HH,VV}}$ reaches only about half the value of the occupations $\rho^\text{2p}_{\text{HH,HH}}$ and $\rho^\text{2p}_{\text{VV,VV}}$. As a result, the degree of entanglement is strongly reduced.

\subsection{Two-photon~$U\vert M$ and two-photon~$N\vert L$ resonance}
\label{subsubsec:2p_UM}

\begin{figure}
\centering
\includegraphics[width=\columnwidth]{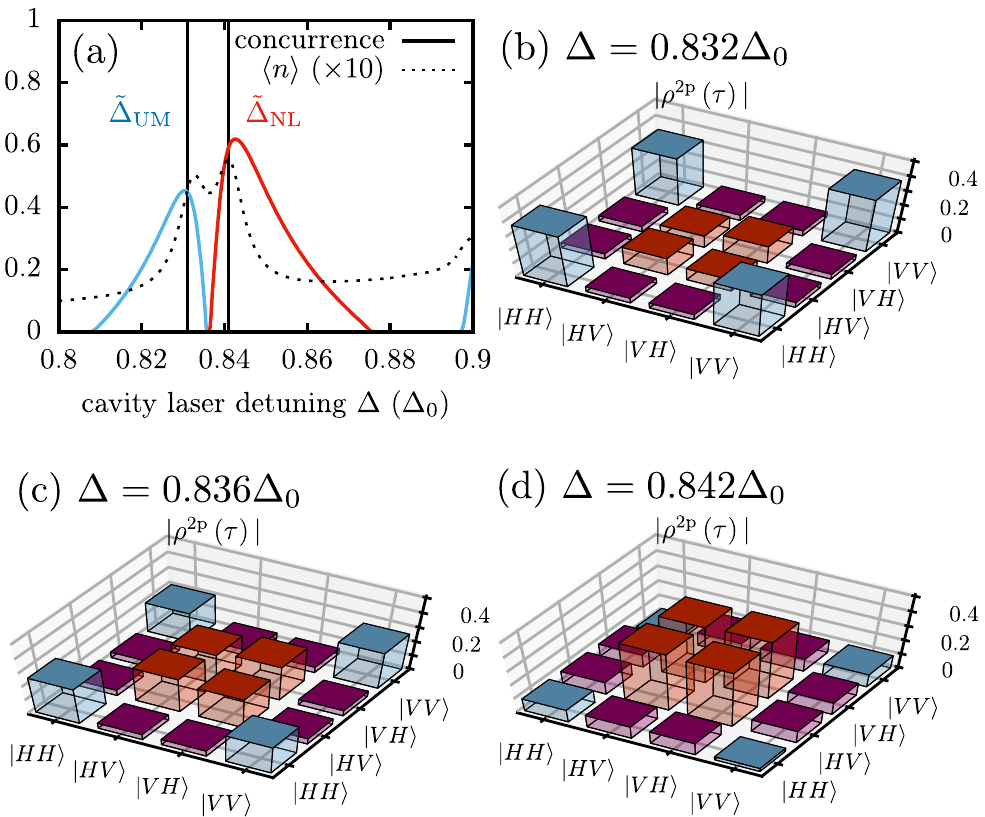}
\caption{(a) Concurrence and mean photon number $\langle n\rangle$ for $\Omega = 30g$. Vertical lines indicate the position of $\tilde{\Delta}_{\text{UM}}$ (Eq.~\eqref{eq:tD_UM}) and $\tilde{\Delta}_{\text{NL}}$ (Eq.~\eqref{eq:tD_NL}). (b)-(d) Absolute values of the two-photon density matrices $\vert\rho^\text{2p}(\tau)\vert$ for $\Delta$ as indicated.}
\label{fig:zoom_30g}
\end{figure}

Next we consider the two-photon resonances between the laser-dressed states $\KET{U}$ and $\KET{M}$, and between $\KET{N}$ and $\KET{L}$, which have the same energy. In Fig.~\ref{fig:concurrence_vs_detuning}, these resonances are indicated by a dark green line. From Fig.~\ref{fig:concurrence_vs_detuning}, we see that here always a sharp transition between $\Phi$BS and $\Psi$BS entanglement takes place. This is highlighted in Fig.~\ref{fig:zoom_30g}(a), which presents a closer look at this resonance condition for $\Omega=30g$. Figure~\ref{fig:zoom_30g}(b)-(d) display the corresponding two-photon density matrices for three selected detuning values. With rising cavity laser detuning the entangled state created inside the cavity changes from $\Phi$BS to $\Psi$BS entanglement, passing trough a special point at $\Delta\approx0.836\Delta_0$ where the concurrence drops to zero.

Here, we have two transitions, for which the corresponding Schrieffer-Wolff analysis yields the Hamiltonians
\begin{eqnarray} \label{eq:SWT_UM}
&&\hat{\tilde{H}}_{\text{UM}}^{(2)} 
= g^2
\begin{pmatrix}
\delta_1^{\text{UM}}-\delta_2^{\text{UM}} & 0 & 0 & \gamma_2^{\text{UM}}\\
0 & \delta_3^{\text{UM}} & \alpha^{\text{UM}} & 0\\
0 & \alpha^{\text{UM}} & \delta_3^{\text{UM}} & 0 \\
\gamma_2^{\text{UM}} & 0 & 0 & \delta_3^{\text{UM}}
\end{pmatrix}
\end{eqnarray}
and 
\begin{eqnarray} \label{eq:SWT_NL}
&&\hat{\tilde{H}}_{\text{NL}}^{(2)} 
= g^2
\begin{pmatrix}
\delta_1^{\text{UM}}-\delta_2^{\text{UM}} & \gamma_1^{\text{NL}} &  \gamma_2^{\text{NL}} &0\\
\gamma_1^{\text{NL}} & \delta_3^{\text{NL}} & \alpha^{\text{NL}} & 0\\
\gamma_2^{\text{NL}} & \alpha^{\text{NL}} & \delta_3^{\text{NL}} & 0 \\
0 & 0 & 0 & \delta_3^{\text{NL}}
\end{pmatrix}
\end{eqnarray}
with the coefficients given in App.~\ref{app:subsec:2p_UM}. While the Hamiltonian $\hat{\tilde{H}}_{\text{UM}}^{(2)}$ has the same form as $\hat{\tilde{H}}_{\text{MN}}^{(2)}$ in Eq.~\eqref{eq:result_2p_MN}, the Hamiltonian $\hat{\tilde{H}}_{\text{NL}}^{(2)}$ has a form similar to $\hat{\tilde{H}}_{\text{UL}}^{(2)}$ in Eq.~\eqref{eq:result_2p_UL_v3}. 

From the effective Hamiltonian, it is evident that the isolated 2p~$U\vert M$ resonance supports only $\Phi$BS entanglement, while the isolated 2p~$N\vert L$ resonance has competing channels for both $\Phi$BS and $\Psi$BS entanglement. From the coefficients, we can deduce the strengths of the competing channels, finding that
\begin{equation}
	\ABS{\gamma_1^{\text{NL}}} = \ABS{\gamma_2^{\text{NL}}} + \frac{2\sqrt{2}\tilde{c}}{2\Delta_0+\dcL{UM}}\,.
\end{equation}
Therefore the ratio $\gamma_1^{\text{NL}}/\gamma_2^{\text{NL}}$ is always larger than 1 and the preferred type of entanglement for the 2p~$N\vert L$ resonance is always $\Psi$BS entanglement.

A zoom in around the two-photon transition at $\Delta=\dcL{UM}/2$, presented in Fig.~\ref{fig:zoom_30g}(a) for $\Omega=30g$, shows clearly that two peaks appear, a $\Phi$BS one and a $\Psi$BS one. The approximate position of these peaks can be determined by the diagonal elements of the Schrieffer-Wolff Hamiltonians in Eq.~\eqref{eq:SWT_UM} and Eq.~\eqref{eq:SWT_NL}. Due to the transformation, diagonal elements appear encoded by $\delta_j^{\chi_1\chi_2}$, which slightly shift the resulting resonance, such that now we have the resonances for the 2p~$U\vert M$ transition with $\Phi$BS entanglement at
\begin{eqnarray} \label{eq:tD_UM}
	\tilde{\Delta}_{\text{UM}} = \frac{1}{2}\left(\dcL{UM} + (\delta_1^{\text{UM}}-\delta_2^{\text{UM}})-\delta_3^{\text{UM}}\right)
\end{eqnarray}
and the 2p~$N\vert L$ transition with $\Psi$BS entanglement at
\begin{eqnarray}\label{eq:tD_NL}
	\tilde{\Delta}_{\text{NL}} = \frac{1}{2}\left(\dcL{UM} + (\delta_1^{\text{UM}}-\delta_2^{\text{UM}})-\delta_3^{\text{NL}}\right)\,.
\end{eqnarray}
The values of the different $\delta_j^{\chi_1\chi_2}$ are given in App.~\ref{app:subsec:2p_UM}. Indeed, the position of the peak maxima visible in Fig.~\ref{fig:zoom_30g} agree well with these shifted resonances (indicated by vertical lines). This interpretation is confirmed by the mean photon number $\langle n \rangle$ (dotted line in Fig.~\ref{fig:zoom_30g}) which also displays two separate maxima, indicating two close-by resonances [cf., Fig.~\ref{fig:zoom_30g}(a)]. 

Also, the $\delta_j^{\chi_1\chi_2}$ depend sensibly on the driving strength $\Omega$. For a driving strength being smaller than $\Omega_{\text{m}} = \sqrt{3} \Delta_0 \approx 34.6g$ we find that $\tilde{\Delta}_{\text{UM}} < \tilde{\Delta}_{\text{NL}}$, while for $\Omega>\Omega_{\text{m}}$ this order is reversed. Therefore, in Fig.~\ref{fig:concurrence_vs_detuning}(d) for a driving strength $\Omega = 40g$ the arrangement of $\Psi$BS and $\Phi$BS entanglement is swapped. 

In between the regions of $\Phi$BS and $\Psi$BS entanglement we have the special point at  $(\tilde{\Delta}_{\text{UM}}+ \tilde{\Delta}_{\text{NL}})/2$. From the density matrix at this special point [cf. Fig.~\ref{fig:zoom_30g}(c)], we see that the concurrence does not vanish due to the lack of coherences. We find that at the special point the generated two-photon state is essentially the superposition of the two density matrices created by each transition individually with 
\begin{equation}
\rho^\text{2p}_{\text{sp2}} = \frac{1}{2}\left[
\frac{1}{2}
\begin{pmatrix}
0 & 0 & 0 & 0\\
0 & 1 & 1 & 0\\
0 & 1 & 1 & 0\\
0 & 0 & 0 & 0
\end{pmatrix}
+\frac{1}{2}
\begin{pmatrix}
1 & 0 & 0 & -1\\
0 & 0 & 0 & 0\\
0 & 0 & 0 & 0\\
-1 & 0 & 0 & 1
\end{pmatrix}
\right]
\end{equation}
This can be rewritten into
\begin{equation}
\label{eq:rho_sp_2}
\rho^\text{2p}_{\text{sp2}} = \frac{1}{2}\KET{\psi_{\text{sp2}}^{(+)}}\BRA{\psi_{\text{sp2}}^{(+)}} + \frac{1}{2}\KET{\psi_{\text{sp2}}^{(-)}}\BRA{\psi_{\text{sp2}}^{(-)}},
\end{equation}
with
\begin{equation}
\KET{\psi_{\text{sp2}}^{(\pm)}} = \frac{1}{\sqrt{2}} \left( \KET{H_1} \pm i \KET{V_1} \right) \frac{1}{\sqrt{2}} \left( \KET{H_2} \pm i \KET{V_2} \right)\,.
\end{equation}
Thus, the density matrix can be written as a mixed state, where both contributing states are products of two one-photon states, i.e., the states are factorizable states, and, accordingly, the corresponding concurrence vanishes. 

We emphasize that this is a different type of special point than the one discussed in Sec.~\ref{subsec:transition_del_Valle} where the system approaches a pure factorizable state. Another difference in comparison to the 2p~$U\vert L$ resonance can be found in the limit $\Omega\rightarrow\infty$. While the concurrence obtained at the 2p~$U\vert L$ resonance approaches a high finite value and becomes independent of the driving strength, the concurrence for the 2p~$U\vert M$ and 2p~$N\vert L$ resonances approach zero. In the limiting case the difference $\tilde{\Delta}_\text{UM}-\tilde{\Delta}_\text{NL}$ vanishes and, therefore, the two resonances merge together and the different types of entanglement cancel each other.

\subsection{Two-photon~$U\vert N$ and two-photon~$M\vert L$ resonance}
\label{subsubsec:2p_UN}

Finally, we analyze the remaining two resonances 2p~$U\vert N$ and 2p~$M\vert L$. In Fig.~\ref{fig:concurrence_vs_detuning} we see that always $\Phi$BS occurs at this transition. 

The analysis with the Schrieffer-Wolff transformation results in a similar situation as discussed in the previous subsection \ref{subsubsec:2p_UM}: The Hamiltonian of the 2p~$M\vert L$ transition has the same form as the 2p~$U\vert M$ transition [Eq.~\eqref{eq:SWT_UM} or also Eq.~\eqref{eq:result_2p_MN}] and therefore promotes exclusively $\Phi$BS entanglement. On the other hand, the Hamiltonian of the 2p~$U\vert N$ transition has the same form as the 2p~$N\vert L$ transition [Eq.~\eqref{eq:SWT_NL} or also Eq.~\eqref{eq:result_2p_UL_v3}] and therefore promotes both $\Phi$BS and $\Psi$BS entanglement. The dominating type of entanglement depends on the ratio of $\gamma_1^{\text{UN}}$ to $\gamma_2^{\text{UN}}$, but also on the splitting from the other resonances given by the diagonal elements $\delta_{j}^{\chi_1\chi_2}$. For small driving strength values $\Omega<20g$ the 2p~$U\vert N$ transition dominates the dynamics and the resulting entanglement is $\Phi$BS entanglement. For larger $\Omega$ both two-photon resonances become of equal importance and a transition between $\Phi$BS and $\Psi$BS entanglement is expected, similar to the results presented in Sec.~\ref{subsubsec:2p_UM}. But, in contrast to the previous section, here, the splitting of the two peaks is too small for the given driving strength values, therefore, we only observe $\Phi$BS entanglement in Fig.~\ref{fig:concurrence_vs_detuning}.

The corresponding Hamiltonians and constants are given in App.~\ref{app:subsec:2p_UN}.

\section{Conclusion} 
\label{sec:conclusion}

In conclusion, we have investigated the possible types of entanglement generated by a driven four-level emitter --cavity system. We found that two different types of entanglement can occur, which we classified as $\Phi$BS and $\Psi$BS entanglement. 

By adjusting the driving strength as well as the cavity detuning, we found a rich picture showing a finite concurrence at various transitions.  Using a Schrieffer-Wolff transformation, we were able to give analytical insight into the occurance of the different types of entanglement showing that either $\Phi$BS or a mixture of $\Phi$BS and $\Psi$BS is promoted at the two-photon transitions. Most excitingly, we found special points, where the concurrence, a measure for the entanglement, drops to zero, though the corresponding coherences in the two-photon density matrix are not absent. Instead, factorizable (and therefore not entangled states) are reached. 

Seeing that entanglement, being one of the most remarkable and interesting physical effects that separates the quantum mechanical from the classical world, can change its character by just adding an external driving to a few-level emitter is exciting from a fundamental point of view and can also lead to new possibilities for using few-level emitters in quantum information technology.

\section*{Acknowledgments}

M. Cygorek thanks the Alexander-von-Humboldt foundation for support through a Feodor Lynen fellowship. 
A. Vagov acknowledges the support from the Russian Science Foundation under the Project 18-12-00429 which was used to study dynamical processes leading to two-photon entanglement.
D. E. Reiter acknowledges support by the Deutsche Forschungsgemeinschaft (DFG) via the project 428026575.
We are further greatful for support by the Deutsche
Forschungsgemeinschaft (DFG, German Research Foundation) via the project 419036043.

\renewcommand{\theequation}{A\arabic{equation}}
\renewcommand{\thesection}{A}
\renewcommand{\thesubsection}{\arabic{subsection}}
\setcounter{equation}{0}

\section{Schrieffer-Wolff Transformation}
\label{app:SWT}

For the Schrieffer-Wolff transformation we consider the FLE-cavity system without losses and use the states $\KET{\chi,n_\text{H},n_\text{V}}$ where 
$\KET{\chi}\in\left\lbrace\KET{U},\KET{M},\KET{N},\KET{L}\right\rbrace$ is one of the four laser-dressed states defined in Sec.~\ref{subsec:dressed_state_pic} and $n_\text{H}$ ($n_\text{V}$) denotes the number of photons present in the horizontally (vertically) polarized cavity mode. The direct two-photon transition from $\KET{\chi_1}$ to $\KET{\chi_2}$ involves only the states
\begin{equation}\label{eq:subA}
	A:\KET{\chi_1,0,0},\KET{\chi_2,1,1}, \KET{\chi_2,2,0},\KET{\chi_2,0,2} .
\end{equation}
As discussed in Sec.~\ref{subsec:SWT}, there are also several other paths to create the two-photon states, thereby coupling the aforementioned states. These processes are depicted in Fig.~\ref{fig:SchriefferWolffTrafo} and include the states
\begin{eqnarray} \label{eq:subB}
	B&:&\KET{\chi,1,0}, \KET{\chi,0,1},\notag \\
	&& \KET{\chi,3,0},\KET{\chi,2,1}, \KET{\chi,1,2},\KET{\chi,0,3},
\end{eqnarray}
where the one- and three-photon states include all four bare states, i.e., $\KET{\chi} = \KET{U},\KET{N},\KET{M},\KET{L}$. This results in a $28\times 28$ matrix. To reduce this to a $4\times4$ matrix for the relevant states in subset $A$ [see~\eqref{eq:subA}], we use a Schrieffer-Wolff transformation \cite{Winkler,LossDivincenzo_SW}. In the transformation, we perform a block-diagonalization of the system Hamiltonian via the unitary transformation 
\begin{equation}
	e^{-\hat{S}}\hat{H}e^{\hat{S}}\,,
\end{equation}
where $\hat{S}$ is an anti-Hermitian operator \cite{Winkler}. After the decoupling procedure, the states in set $B$ [see \eqref{eq:subB}] can be disregarded as they are insignificant for the system dynamics. This formalism can be applied here since, for a given two-photon resonance, where the cavity laser detuning matches half the transition energy between the states $\KET{\chi_1}$ and $\KET{\chi_2}$, one-photon transition processes between the laser-dressed states are typically strongly off-resonant.

In second order the effective Hamiltonian for the states in set $A$ is then given by $\hat{H}_{\chi_1\chi_2}^{(2)}= \left\{H_{a,a'}^{(2)}\right\}_{\chi_1\chi_2}$ with the matrix elements \cite{Winkler}
\begin{eqnarray}
\label{app:eq:SWT}
H_{a,a'}^{(2)} &=& H_{a,a'} \\
&&+ \frac{1}{2}\left\lbrace \sum_b H_{a,b} H_{b, a'} \left[\frac{1}{E_{a}-E_{b}}+\frac{1}{E_{a'}-E_{b}} \right] \right\rbrace , \notag
\end{eqnarray}
where $a$ runs over the states in subset $A$, the index $b$ runs over the states in $B$, and
\begin{equation}
	E_j = \BRA{j}\hat{H}\KET{j} = E_{\chi}+\left(n_\text{H}+n_\text{V}\right)\Delta
\end{equation}
is the energy of the state $\KET{j}=\KET{\chi,n_\text{H},n_\text{V}}\in A,B$.
The matrix elements are calculated from the system Hamiltonian with 
\begin{equation}
	H_{a,a^\prime} = E_a\delta_{a,a^\prime}\,,
\end{equation}
This term can be dropped since it represents a constant energy shift as the four states in set $A$ are energetically degenerate. The remaining matrix elements for $a \neq b$ are given by the coupling Hamiltonian in the dressed state basis [Eq.~\eqref{eq:H_DS}] with
\begin{equation}
	H_{a,b} = \BRA{a} \hat{H}_{\text{DS-c}} \KET{b}
\end{equation}
After the Schrieffer-Wolff transformation we perform a rotation to the basis
\begin{eqnarray}
	\KET{\chi_1,0,0},\KET{\chi_2,1,1}, \KET{\chi_2,\Phi_+},\KET{\chi_2,\Phi_-}\,
\end{eqnarray}
using 
\begin{eqnarray}
	\hat{\tilde{H}}_{\chi_1\chi_2}^{(2)} &=& T^{\dagger}\hat{H}^{(2)}_{\chi_1\chi_2}T
	\,\, \text{with} \,\, T =\begin{pmatrix}
		1&0&0&0\\
		0&1&0&0\\
		0&0&\frac{1}{\sqrt{2}}&\frac{1}{\sqrt{2}}\\
		0&0&\frac{1}{\sqrt{2}}&-\frac{1}{\sqrt{2}}\\		
		\end{pmatrix}
\end{eqnarray}
We performed this procedure for all two-photon resonances. 

\subsection{Effective Hamiltonian for the 2p~$U\vert L$ resonance}
\label{app:subsec:2p_UL}

The effective Hamiltonian is
\begin{eqnarray}
&&\hat{\tilde{H}}_{\text{UL}}^{(2)} 
= g ^2 \\
&&\times \begin{pmatrix}
\delta^{\text{UL}} & \gamma_1^{\text{UL}} & -\gamma_2^{\text{UL}} & 0\\
\gamma_1^{\text{UL}} & -\delta^{\text{UL}}-\delta_3^\text{UL} & \alpha^{\text{UL}} & 0\\
-\gamma_2^{\text{UL}} & \alpha^{\text{UL}} & -\delta^{\text{UL}}-\delta_3^\text{UL} & 0\\
0 & 0 & 0 & -\delta^{\text{UL}}-\delta_3^\text{UL}
\end{pmatrix} \notag
\end{eqnarray}
in the basis $\KET{U,0,0}$, $\KET{L,1,1}$, $\KET{L,\Phi_+}$ and $\KET{L,\Phi_-}$ with
\begin{eqnarray}
\delta^{\text{UL}} &=& \left(\tilde{c}^2-c^2\right) \left(\frac{2}{\Delta_0}+\frac{4}{\dcL{UL}}\right) \notag \\
\delta_3^\text{UL} &=& \frac{8\left(\tilde{c}^2-c^2\right)^2}{3\Delta_\text{UL}} + \frac{2\tilde{c}^2}{\Delta_\text{UL}+\Delta_0/2} +\frac{2c^2}{\Delta_\text{UL}-\Delta_0/2} \notag\\
\gamma_1^{\text{UL}} &=& 4c \tilde{c}\frac{1}{\Delta_0}  - 16 c\tilde{c}\left(\tilde{c}^2- c^2\right)\frac{1}{\dcL{UL}} \notag \\ 
\gamma_2^{\text{UL}} &=& 16 c\tilde{c}\left(\tilde{c}^2- c^2\right)\frac{1}{\dcL{UL}} \notag \\
\alpha^{\text{UL}} &=& \frac{1}{\Delta_0} - \left(1 -16 c^2 \tilde{c}^2\right)\frac{1}{\dcL{UL}} -\frac{1}{2}\delta_3^\text{UL} +\frac{2\tilde{c}^2}{\Delta_\text{UL}+\Delta_0/2}  \,. \notag 
\end{eqnarray}

\subsection{Effective Hamiltonian for the 2p~$M\vert N$ resonance}
\label{app:subsec:2p_MN}
The effective Hamiltonian is
\begin{eqnarray}
&&\hat{\tilde{H}}_{\text{MN}}^{(2)} 
= g^2 \\
&&\times\begin{pmatrix}
\delta^{\text{MN}} & 0 & 0 & \gamma_2^{\text{MN}}\\
0 & -\delta^{\text{MN}}+\delta_3^\text{MN} & \alpha^\text{MN} & 0\\
0 & \alpha^\text{MN} & -\delta^{\text{MN}}+\delta_3^\text{MN} & 0 \\
\gamma_2^{\text{MN}} & 0 & 0 & -\delta^{\text{MN}}+\delta_3^\text{MN}
\end{pmatrix} \notag
\end{eqnarray}
in the basis $\KET{M,0,0}$, $\KET{N,1,1}$, $\KET{N,\Phi_+}$ and $\KET{N,\Phi_-}$ with
\begin{eqnarray}
\delta^{\text{MN}} &=& 2\left(\tilde{c}^2-c^2\right)\frac{1}{\dcL{UL}} \notag \\
\delta_3^\text{MN} &=& -\frac{4\tilde{c}^2}{2\Delta_0+\Delta_\text{UL}} -\frac{2}{3\Delta_0} -\frac{4c^2}{2\Delta_0-\Delta_\text{UL}}\notag \\ 
\gamma_2^{\text{MN}} &=& -4\,c\,\tilde{c}\frac{1}{\dcL{UL}} \notag \\
\alpha^\text{MN} &=& -\delta^\text{MN} +\frac{1}{2}\delta_3^\text{MN} + \frac{1}{3\Delta_0}. \notag 
\end{eqnarray}

\subsection{Effective Hamiltonians for the 2p~$U\vert M$ and 2p~$N\vert L$ resonance}
\label{app:subsec:2p_UM}

The effective Hamiltonian for the 2p~$U\vert M$ resonance is
\begin{eqnarray}
&&\hat{\tilde{H}}_{\text{UM}}^{(2)} 
= g^2\\
&&\times \begin{pmatrix}
\delta_1^{\text{UM}}-\delta_2^{\text{UM}} & 0 & 0 & \gamma_2^{\text{UM}}\\
0 & \delta_3^{\text{UM}} & \alpha^{\text{UM}} & 0\\
0 & \alpha^{\text{UM}} & \delta_3^{\text{UM}} & 0 \\
\gamma_2^{\text{UM}} & 0 & 0 & \delta_3^{\text{UM}}
\end{pmatrix}
 \text{for}\,
\begin{array}{l} \KET{U,0,0}\\ \KET{M,1,1} \\ \KET{M,\Phi_+}\\ \KET{M,\Phi_-}\end{array} \notag 
\end{eqnarray}
where
\begin{subequations}
\begin{eqnarray}
\delta_1^{\text{UM}} &=& -\frac{16c^2\tilde{c}^2}{\dcL{UM}} + \frac{2\tilde{c}^2}{2\Delta_0+\dcL{UM}} \notag 
+ \frac{4(\tilde{c}^2-c^2)^2}{2\Delta_0+3\dcL{UM}} \notag \\
\delta_2^{\text{UM}} &=&  -\frac{2c^2}{\dcL{UM}} + \frac{1}{2\Delta_0+\dcL{UM}} + \frac{2\tilde{c}^2}{2\Delta_0+3\dcL{UM}} \notag \\
\delta_3^{\text{UM}} &=& -\frac{4c^2}{3\dcL{UM}} +\frac{2}{2\Delta_0-\dcL{UM}} +\frac{4\tilde{c}^2}{2\Delta_0+\dcL{UM}} \notag \\
\gamma_2^{\text{UM}} &=& -\frac{4\sqrt{2}c^2\tilde{c}}{\dcL{UM}} - \frac{\sqrt{2}\tilde{c}}{2\Delta_0+\dcL{UM}} 
	+ \frac{2\sqrt{2}\left(\tilde{c}^2-c^2\right)\tilde{c}}{2\Delta_0+3\dcL{UM}} \notag \\
\alpha^{\text{UM}} &=& - \delta_2^{\text{UM}} -\frac{1}{2}\delta_3^{\text{UM}} \notag 
\end{eqnarray}
\end{subequations}
The effective Hamiltonian for the two-photon transition between the states $\KET{N}$ and $\KET{L}$ is given by 
\begin{eqnarray}
&&\hat{\tilde{H}}_{\text{NL}}^{(2)} 
= g^2\\
&&\times \begin{pmatrix}
\delta_1^{\text{UM}}-\delta_2^{\text{UM}} & \gamma_1^{\text{NL}} &  \gamma_2^{\text{NL}} &0\\
\gamma_1^{\text{NL}} & \delta_3^{\text{NL}} & \alpha^{\text{NL}} & 0\\
\gamma_2^{\text{NL}} & \alpha^{\text{NL}} & \delta_3^{\text{NL}} & 0 \\
0 & 0 & 0 & \delta_3^{\text{NL}}
\end{pmatrix}
 \text{for}\,
\begin{array}{l} \KET{N,0,0}\\ \KET{L,1,1} \\ \KET{L,\Phi_+}\\ \KET{L,\Phi_-}\end{array} \notag 
\end{eqnarray}
with
\begin{subequations}
\begin{eqnarray}
\delta_3^{\text{NL}} &=&  - \frac{32c^2\tilde{c}^2}{\dcL{UM}}-\frac{4c^2}{3\dcL{UM}} -\frac{8(\tilde{c}^2-c^2)}{2\Delta_0+5\dcL{UM}} -\frac{4\tilde{c}^2}{2\Delta_0+3\dcL{\text{UM}}}  \notag \\
\gamma_1^{\text{NL}} &=& \gamma_2^{\text{UM}}   \notag\\
\gamma_2^{\text{NL}} &=& \gamma_2^{\text{UM}} + \frac{2\sqrt{2}\tilde{c}}{2\Delta_0+\dcL{UM}}  \notag\\
\alpha^{\text{NL}} &=& -\delta_1^{\text{UM}}+\frac{1}{2}\delta_3^{\text{NL}}
	+\frac{4\tilde{c}^2}{2\Delta_0+\dcL{UM}} +\frac{4\tilde{c}^2}{2\Delta_0+3\dcL{UM}} \notag
\end{eqnarray}
\end{subequations}

\subsection{Effective Hamiltonians for the 2p~$U\vert N$ and 2p~$M\vert L$ resonance}
\label{app:subsec:2p_UN}

For the 2p~$U\vert N$ transition we obtain
\begin{eqnarray}
&&\hat{\tilde{H}}_{\text{UN}}^{(2)} 
= g^2\\
&&\times
\begin{pmatrix}
\delta_1^{\text{UN}}-\delta_2^{\text{UN}} & \gamma_1^{\text{UN}} &\gamma_2^{\text{UN}} & 0\\
\gamma_1^{\text{UN}} & \delta_3^{\text{UN}} & \alpha^{\text{UN}} & 0\\
\gamma_2^{\text{UN}} & \alpha^{\text{UN}} & \delta_3^{\text{UN}} &  0\\
0 & 0 & 0 & \delta_3^{\text{UN}}
\end{pmatrix}
\text{for}\,
\begin{array}{l} \KET{U,0,0}\\ \KET{N,1,1} \\ \KET{N,\Phi_+}\\ \KET{N,\Phi_-}\end{array} \notag .
\end{eqnarray}
The energies and coupling strengths are

\begin{eqnarray}
\delta_1^{\text{UN}} &=& -\frac{16c^2\tilde{c}^2}{\dcL{UN}} + \frac{2c^2}{\dcL{UN}-2\Delta_0}   
	+ \frac{4\left(\tilde{c}^2-c^2\right)^2}{3\dcL{UN}-2\Delta_0}\notag \\
\delta_2^{\text{UN}} &=& -\frac{2\tilde{c}^2}{\dcL{UN}} + \frac{1}{\dcL{UN}-2\Delta_0} 
	+\frac{2c^2}{3\dcL{UN}-2\Delta_0} \notag \\
\delta_3^{\text{UN}} &=& -\frac{4\tilde{c}^2}{3\dcL{UN}} -\frac{2}{2\Delta_0+\dcL{UN}} 
	-\frac{4c^2}{2\Delta_0-\dcL{UN}} \notag \\
\gamma_1^{\text{UN}} &=& -\frac{4\sqrt{2} c \tilde{c}^2}{\dcL{UN}} - \frac{\sqrt{2}c}{\dcL{UN}-2\Delta_0}
	- \frac{2\sqrt{2}\left(\tilde{c}^2- c^2\right)c}{3\dcL{UN}-2\Delta_0}\notag \\
\gamma_2^{\text{UN}} &=&  \gamma_1^{\text{UN}}+ \frac{2\sqrt{2}c}{\dcL{UN}-\Delta_0} \notag \\
\alpha^{\text{UN}} &=& \delta_2^{\text{UN}}+\frac{1}{2}\delta_3^{\text{UN}}
	-\frac{2}{\dcL{UN}-2\Delta_0} 
	+\frac{2}{2\Delta_0+\dcL{UN}} \notag .
\end{eqnarray}

For the 2p~$M\vert L$ transition we have
\begin{eqnarray}
&&\hat{\tilde{H}}_{\text{ML}}^{(2)} 
= g^2\\
\times &&\begin{pmatrix}
\delta_1^{\text{UN}}-\delta_2^{\text{UN}} & 0 & 0 & \gamma_1^{\text{UN}}\\
0 & \delta_3^{\text{ML}} & \alpha^{\text{ML}} & 0\\
0 & \alpha^{\text{ML}} & \delta_3^{\text{ML}} & 0 \\
\gamma_1^{\text{UN}} & 0 & 0 & \delta_3^{\text{ML}}
\end{pmatrix}
\text{for}\,
\begin{array}{l} \KET{M,0,0}\\ \KET{L,1,1} \\ \KET{L,\Phi_+}\\ \KET{L,\Phi_-}\end{array} \notag 
\end{eqnarray}
with
\begin{eqnarray}
\delta_3^{\text{ML}} &=&\frac{ 8\left(\tilde{c}^2- c^2\right)^2}{2\Delta_0-5\dcL{UN}} -\frac{4\tilde{c}^2}{3\dcL{UN}} 
	+ \frac{4c^2}{2\Delta_0-3\dcL{UN}} -\frac{32 c^2 \tilde{c}^2}{\dcL{UN}} \notag \\
\alpha^{\text{ML}} &=& -\delta_1^{\text{UN}} +\frac{1}{2}\delta_3^{\text{ML}}+\frac{4\tilde{c}^2}{3\dcL{UN}}  \notag .
\end{eqnarray}

\bibliography{PIbib}

\begin{thebibliography}{57}%
\makeatletter
\providecommand \@ifxundefined [1]{%
 \@ifx{#1\undefined}
}%
\providecommand \@ifnum [1]{%
 \ifnum #1\expandafter \@firstoftwo
 \else \expandafter \@secondoftwo
 \fi
}%
\providecommand \@ifx [1]{%
 \ifx #1\expandafter \@firstoftwo
 \else \expandafter \@secondoftwo
 \fi
}%
\providecommand \natexlab [1]{#1}%
\providecommand \enquote  [1]{``#1''}%
\providecommand \bibnamefont  [1]{#1}%
\providecommand \bibfnamefont [1]{#1}%
\providecommand \citenamefont [1]{#1}%
\providecommand \href@noop [0]{\@secondoftwo}%
\providecommand \href [0]{\begingroup \@sanitize@url \@href}%
\providecommand \@href[1]{\@@startlink{#1}\@@href}%
\providecommand \@@href[1]{\endgroup#1\@@endlink}%
\providecommand \@sanitize@url [0]{\catcode `\\12\catcode `\$12\catcode
  `\&12\catcode `\#12\catcode `\^12\catcode `\_12\catcode `\%12\relax}%
\providecommand \@@startlink[1]{}%
\providecommand \@@endlink[0]{}%
\providecommand \url  [0]{\begingroup\@sanitize@url \@url }%
\providecommand \@url [1]{\endgroup\@href {#1}{\urlprefix }}%
\providecommand \urlprefix  [0]{URL }%
\providecommand \Eprint [0]{\href }%
\providecommand \doibase [0]{http://dx.doi.org/}%
\providecommand \selectlanguage [0]{\@gobble}%
\providecommand \bibinfo  [0]{\@secondoftwo}%
\providecommand \bibfield  [0]{\@secondoftwo}%
\providecommand \translation [1]{[#1]}%
\providecommand \BibitemOpen [0]{}%
\providecommand \bibitemStop [0]{}%
\providecommand \bibitemNoStop [0]{.\EOS\space}%
\providecommand \EOS [0]{\spacefactor3000\relax}%
\providecommand \BibitemShut  [1]{\csname bibitem#1\endcsname}%
\let\auto@bib@innerbib\@empty
\bibitem [{\citenamefont {Horodecki}\ \emph {et~al.}(2009)\citenamefont
  {Horodecki}, \citenamefont {Horodecki}, \citenamefont {Horodecki},\ and\
  \citenamefont {Horodecki}}]{Horodecki:09}%
  \BibitemOpen
  \bibfield  {author} {\bibinfo {author} {\bibfnamefont {R.}~\bibnamefont
  {Horodecki}}, \bibinfo {author} {\bibfnamefont {P.}~\bibnamefont
  {Horodecki}}, \bibinfo {author} {\bibfnamefont {M.}~\bibnamefont
  {Horodecki}}, \ and\ \bibinfo {author} {\bibfnamefont {K.}~\bibnamefont
  {Horodecki}},\ }\href {\doibase 10.1103/RevModPhys.81.865} {\bibfield
  {journal} {\bibinfo  {journal} {Rev. Mod. Phys.}\ }\textbf {\bibinfo {volume}
  {81}},\ \bibinfo {pages} {865} (\bibinfo {year} {2009})}\BibitemShut
  {NoStop}%
\bibitem [{\citenamefont {Orieux}\ \emph {et~al.}(2017)\citenamefont {Orieux},
  \citenamefont {Versteegh}, \citenamefont {J{\"o}ns},\ and\ \citenamefont
  {Ducci}}]{Orieux_entangled}%
  \BibitemOpen
  \bibfield  {author} {\bibinfo {author} {\bibfnamefont {A.}~\bibnamefont
  {Orieux}}, \bibinfo {author} {\bibfnamefont {M.~A.~M.}\ \bibnamefont
  {Versteegh}}, \bibinfo {author} {\bibfnamefont {K.~D.}\ \bibnamefont
  {J{\"o}ns}}, \ and\ \bibinfo {author} {\bibfnamefont {S.}~\bibnamefont
  {Ducci}},\ }\href {http://stacks.iop.org/0034-4885/80/i=7/a=076001}
  {\bibfield  {journal} {\bibinfo  {journal} {Rep. Prog. Phys.}\ }\textbf
  {\bibinfo {volume} {80}},\ \bibinfo {pages} {076001} (\bibinfo {year}
  {2017})}\BibitemShut {NoStop}%
\bibitem [{\citenamefont {Akopian}\ \emph {et~al.}(2006)\citenamefont
  {Akopian}, \citenamefont {Lindner}, \citenamefont {Poem}, \citenamefont
  {Berlatzky}, \citenamefont {Avron}, \citenamefont {Gershoni}, \citenamefont
  {Gerardot},\ and\ \citenamefont {Petroff}}]{entangled-photon2}%
  \BibitemOpen
  \bibfield  {author} {\bibinfo {author} {\bibfnamefont {N.}~\bibnamefont
  {Akopian}}, \bibinfo {author} {\bibfnamefont {N.~H.}\ \bibnamefont
  {Lindner}}, \bibinfo {author} {\bibfnamefont {E.}~\bibnamefont {Poem}},
  \bibinfo {author} {\bibfnamefont {Y.}~\bibnamefont {Berlatzky}}, \bibinfo
  {author} {\bibfnamefont {J.}~\bibnamefont {Avron}}, \bibinfo {author}
  {\bibfnamefont {D.}~\bibnamefont {Gershoni}}, \bibinfo {author}
  {\bibfnamefont {B.~D.}\ \bibnamefont {Gerardot}}, \ and\ \bibinfo {author}
  {\bibfnamefont {P.~M.}\ \bibnamefont {Petroff}},\ }\href {\doibase
  10.1103/PhysRevLett.96.130501} {\bibfield  {journal} {\bibinfo  {journal}
  {Phys. Rev. Lett.}\ }\textbf {\bibinfo {volume} {96}},\ \bibinfo {pages}
  {130501} (\bibinfo {year} {2006})}\BibitemShut {NoStop}%
\bibitem [{\citenamefont {Gisin}\ \emph {et~al.}(2002)\citenamefont {Gisin},
  \citenamefont {Ribordy}, \citenamefont {Tittel},\ and\ \citenamefont
  {Zbinden}}]{Gisin:02}%
  \BibitemOpen
  \bibfield  {author} {\bibinfo {author} {\bibfnamefont {N.}~\bibnamefont
  {Gisin}}, \bibinfo {author} {\bibfnamefont {G.}~\bibnamefont {Ribordy}},
  \bibinfo {author} {\bibfnamefont {W.}~\bibnamefont {Tittel}}, \ and\ \bibinfo
  {author} {\bibfnamefont {H.}~\bibnamefont {Zbinden}},\ }\href {\doibase
  10.1103/RevModPhys.74.145} {\bibfield  {journal} {\bibinfo  {journal} {Rev.
  Mod. Phys.}\ }\textbf {\bibinfo {volume} {74}},\ \bibinfo {pages} {145}
  (\bibinfo {year} {2002})}\BibitemShut {NoStop}%
\bibitem [{\citenamefont {Lo}\ \emph {et~al.}(2014)\citenamefont {Lo},
  \citenamefont {Curty},\ and\ \citenamefont
  {Tamaki}}]{Lo_quantum_cryptography}%
  \BibitemOpen
  \bibfield  {author} {\bibinfo {author} {\bibfnamefont {H.-K.}\ \bibnamefont
  {Lo}}, \bibinfo {author} {\bibfnamefont {M.}~\bibnamefont {Curty}}, \ and\
  \bibinfo {author} {\bibfnamefont {K.}~\bibnamefont {Tamaki}},\ }\href
  {\doibase 10.1038/nphoton.2014.149} {\bibfield  {journal} {\bibinfo
  {journal} {Nat. Photon.}\ }\textbf {\bibinfo {volume} {8}},\ \bibinfo {pages}
  {595} (\bibinfo {year} {2014})}\BibitemShut {NoStop}%
\bibitem [{\citenamefont {Duan}\ \emph {et~al.}(2001)\citenamefont {Duan},
  \citenamefont {Lukin}, \citenamefont {Cirac},\ and\ \citenamefont
  {Zoller}}]{duan_quantum_comm}%
  \BibitemOpen
  \bibfield  {author} {\bibinfo {author} {\bibfnamefont {L.-M.}\ \bibnamefont
  {Duan}}, \bibinfo {author} {\bibfnamefont {M.~D.}\ \bibnamefont {Lukin}},
  \bibinfo {author} {\bibfnamefont {J.~I.}\ \bibnamefont {Cirac}}, \ and\
  \bibinfo {author} {\bibfnamefont {P.}~\bibnamefont {Zoller}},\ }\href
  {\doibase 10.1038/35106500} {\bibfield  {journal} {\bibinfo  {journal}
  {Nature}\ }\textbf {\bibinfo {volume} {414}},\ \bibinfo {pages} {413}
  (\bibinfo {year} {2001})}\BibitemShut {NoStop}%
\bibitem [{\citenamefont {Huber}\ \emph
  {et~al.}(2018{\natexlab{a}})\citenamefont {Huber}, \citenamefont {Reindl},
  \citenamefont {Aberl}, \citenamefont {Rastelli},\ and\ \citenamefont
  {Trotta}}]{Huber_overview_2018}%
  \BibitemOpen
  \bibfield  {author} {\bibinfo {author} {\bibfnamefont {D.}~\bibnamefont
  {Huber}}, \bibinfo {author} {\bibfnamefont {M.}~\bibnamefont {Reindl}},
  \bibinfo {author} {\bibfnamefont {J.}~\bibnamefont {Aberl}}, \bibinfo
  {author} {\bibfnamefont {A.}~\bibnamefont {Rastelli}}, \ and\ \bibinfo
  {author} {\bibfnamefont {R.}~\bibnamefont {Trotta}},\ }\href {\doibase
  10.1088/2040-8986/aac4c4} {\bibfield  {journal} {\bibinfo  {journal} {Journal
  of Optics}\ }\textbf {\bibinfo {volume} {20}},\ \bibinfo {pages} {073002}
  (\bibinfo {year} {2018}{\natexlab{a}})}\BibitemShut {NoStop}%
\bibitem [{\citenamefont {Pan}\ \emph {et~al.}(2012)\citenamefont {Pan},
  \citenamefont {Chen}, \citenamefont {Lu}, \citenamefont {Weinfurter},
  \citenamefont {Zeilinger},\ and\ \citenamefont {\ifmmode~\dot{Z}\else
  \.{Z}\fi{}ukowski}}]{pan:12}%
  \BibitemOpen
  \bibfield  {author} {\bibinfo {author} {\bibfnamefont {J.-W.}\ \bibnamefont
  {Pan}}, \bibinfo {author} {\bibfnamefont {Z.-B.}\ \bibnamefont {Chen}},
  \bibinfo {author} {\bibfnamefont {C.-Y.}\ \bibnamefont {Lu}}, \bibinfo
  {author} {\bibfnamefont {H.}~\bibnamefont {Weinfurter}}, \bibinfo {author}
  {\bibfnamefont {A.}~\bibnamefont {Zeilinger}}, \ and\ \bibinfo {author}
  {\bibfnamefont {M.}~\bibnamefont {\ifmmode~\dot{Z}\else \.{Z}\fi{}ukowski}},\
  }\href {\doibase 10.1103/RevModPhys.84.777} {\bibfield  {journal} {\bibinfo
  {journal} {Rev. Mod. Phys.}\ }\textbf {\bibinfo {volume} {84}},\ \bibinfo
  {pages} {777} (\bibinfo {year} {2012})}\BibitemShut {NoStop}%
\bibitem [{\citenamefont {Bennett}\ and\ \citenamefont
  {DiVincenzo}(2000)}]{Bennett:00}%
  \BibitemOpen
  \bibfield  {author} {\bibinfo {author} {\bibfnamefont {C.~H.}\ \bibnamefont
  {Bennett}}\ and\ \bibinfo {author} {\bibfnamefont {D.~P.}\ \bibnamefont
  {DiVincenzo}},\ }\href {https://doi.org/10.1038/35005001} {\bibfield
  {journal} {\bibinfo  {journal} {Nature}\ }\textbf {\bibinfo {volume} {404}},\
  \bibinfo {pages} {247} (\bibinfo {year} {2000})}\BibitemShut {NoStop}%
\bibitem [{\citenamefont {Kuhn}\ \emph {et~al.}(2016)\citenamefont {Kuhn},
  \citenamefont {Knorr}, \citenamefont {Reitzenstein},\ and\ \citenamefont
  {Richter}}]{Kuhn:16}%
  \BibitemOpen
  \bibfield  {author} {\bibinfo {author} {\bibfnamefont {S.~C.}\ \bibnamefont
  {Kuhn}}, \bibinfo {author} {\bibfnamefont {A.}~\bibnamefont {Knorr}},
  \bibinfo {author} {\bibfnamefont {S.}~\bibnamefont {Reitzenstein}}, \ and\
  \bibinfo {author} {\bibfnamefont {M.}~\bibnamefont {Richter}},\ }\href
  {\doibase 10.1364/OE.24.025446} {\bibfield  {journal} {\bibinfo  {journal}
  {Opt. Express}\ }\textbf {\bibinfo {volume} {24}},\ \bibinfo {pages} {25446}
  (\bibinfo {year} {2016})}\BibitemShut {NoStop}%
\bibitem [{\citenamefont {Zeilinger}(2017)}]{Zeilinger_entangled}%
  \BibitemOpen
  \bibfield  {author} {\bibinfo {author} {\bibfnamefont {A.}~\bibnamefont
  {Zeilinger}},\ }\href {http://stacks.iop.org/1402-4896/92/i=7/a=072501}
  {\bibfield  {journal} {\bibinfo  {journal} {Physica Scripta}\ }\textbf
  {\bibinfo {volume} {92}},\ \bibinfo {pages} {072501} (\bibinfo {year}
  {2017})}\BibitemShut {NoStop}%
\bibitem [{\citenamefont {Edamatsu}(2007)}]{edamatsu2007entangled}%
  \BibitemOpen
  \bibfield  {author} {\bibinfo {author} {\bibfnamefont {K.}~\bibnamefont
  {Edamatsu}},\ }\href@noop {} {\bibfield  {journal} {\bibinfo  {journal}
  {Japanese Journal of Applied Physics}\ }\textbf {\bibinfo {volume} {46}},\
  \bibinfo {pages} {7175} (\bibinfo {year} {2007})}\BibitemShut {NoStop}%
\bibitem [{\citenamefont {M{\"u}ller}\ \emph {et~al.}(2014)\citenamefont
  {M{\"u}ller}, \citenamefont {Bounouar}, \citenamefont {J{\"o}ns},
  \citenamefont {Gl{\"a}ssl},\ and\ \citenamefont
  {Michler}}]{entangled-photon1}%
  \BibitemOpen
  \bibfield  {author} {\bibinfo {author} {\bibfnamefont {M.}~\bibnamefont
  {M{\"u}ller}}, \bibinfo {author} {\bibfnamefont {S.}~\bibnamefont
  {Bounouar}}, \bibinfo {author} {\bibfnamefont {K.~D.}\ \bibnamefont
  {J{\"o}ns}}, \bibinfo {author} {\bibfnamefont {M.}~\bibnamefont
  {Gl{\"a}ssl}}, \ and\ \bibinfo {author} {\bibfnamefont {P.}~\bibnamefont
  {Michler}},\ }\href {https://doi.org/10.1038/nphoton.2013.377} {\bibfield
  {journal} {\bibinfo  {journal} {Nat. Photon.}\ }\textbf {\bibinfo {volume}
  {8}},\ \bibinfo {pages} {224} (\bibinfo {year} {2014})}\BibitemShut {NoStop}%
\bibitem [{\citenamefont {Hanschke}\ \emph {et~al.}(2018)\citenamefont
  {Hanschke}, \citenamefont {Fischer}, \citenamefont {Appel}, \citenamefont
  {Lukin}, \citenamefont {Wierzbowski}, \citenamefont {Sun}, \citenamefont
  {Trivedi}, \citenamefont {Vuckovi{\'c}}, \citenamefont {Finley},\ and\
  \citenamefont {M{\"u}ller}}]{hanschke2018}%
  \BibitemOpen
  \bibfield  {author} {\bibinfo {author} {\bibfnamefont {L.}~\bibnamefont
  {Hanschke}}, \bibinfo {author} {\bibfnamefont {K.~A.}\ \bibnamefont
  {Fischer}}, \bibinfo {author} {\bibfnamefont {S.}~\bibnamefont {Appel}},
  \bibinfo {author} {\bibfnamefont {D.}~\bibnamefont {Lukin}}, \bibinfo
  {author} {\bibfnamefont {J.}~\bibnamefont {Wierzbowski}}, \bibinfo {author}
  {\bibfnamefont {S.}~\bibnamefont {Sun}}, \bibinfo {author} {\bibfnamefont
  {R.}~\bibnamefont {Trivedi}}, \bibinfo {author} {\bibfnamefont
  {J.}~\bibnamefont {Vuckovi{\'c}}}, \bibinfo {author} {\bibfnamefont {J.~J.}\
  \bibnamefont {Finley}}, \ and\ \bibinfo {author} {\bibfnamefont
  {K.}~\bibnamefont {M{\"u}ller}},\ }\href {\doibase 10.1038/s41534-018-0092-0}
  {\bibfield  {journal} {\bibinfo  {journal} {npj Quantum Inf.}\ }\textbf
  {\bibinfo {volume} {4}},\ \bibinfo {pages} {43} (\bibinfo {year}
  {2018})}\BibitemShut {NoStop}%
\bibitem [{\citenamefont {Huber}\ \emph {et~al.}(2017)\citenamefont {Huber},
  \citenamefont {Reindl}, \citenamefont {Huo}, \citenamefont {Huang},
  \citenamefont {Wildmann}, \citenamefont {Schmidt}, \citenamefont {Rastelli},\
  and\ \citenamefont {Trotta}}]{huber2017}%
  \BibitemOpen
  \bibfield  {author} {\bibinfo {author} {\bibfnamefont {D.}~\bibnamefont
  {Huber}}, \bibinfo {author} {\bibfnamefont {M.}~\bibnamefont {Reindl}},
  \bibinfo {author} {\bibfnamefont {Y.}~\bibnamefont {Huo}}, \bibinfo {author}
  {\bibfnamefont {H.}~\bibnamefont {Huang}}, \bibinfo {author} {\bibfnamefont
  {J.~S.}\ \bibnamefont {Wildmann}}, \bibinfo {author} {\bibfnamefont {O.~G.}\
  \bibnamefont {Schmidt}}, \bibinfo {author} {\bibfnamefont {A.}~\bibnamefont
  {Rastelli}}, \ and\ \bibinfo {author} {\bibfnamefont {R.}~\bibnamefont
  {Trotta}},\ }\href {\doibase 10.1038/ncomms15506} {\bibfield  {journal}
  {\bibinfo  {journal} {Nature Communications}\ }\textbf {\bibinfo {volume}
  {8}},\ \bibinfo {pages} {15506} (\bibinfo {year} {2017})}\BibitemShut
  {NoStop}%
\bibitem [{\citenamefont {Reindl}\ \emph {et~al.}(2017)\citenamefont {Reindl},
  \citenamefont {J\"ons}, \citenamefont {Huber}, \citenamefont {Schimpf},
  \citenamefont {Huo}, \citenamefont {Zwiller}, \citenamefont {Rastelli},\ and\
  \citenamefont {Trotta}}]{reindl2017}%
  \BibitemOpen
  \bibfield  {author} {\bibinfo {author} {\bibfnamefont {M.}~\bibnamefont
  {Reindl}}, \bibinfo {author} {\bibfnamefont {K.~D.}\ \bibnamefont {J\"ons}},
  \bibinfo {author} {\bibfnamefont {D.}~\bibnamefont {Huber}}, \bibinfo
  {author} {\bibfnamefont {C.}~\bibnamefont {Schimpf}}, \bibinfo {author}
  {\bibfnamefont {Y.}~\bibnamefont {Huo}}, \bibinfo {author} {\bibfnamefont
  {V.}~\bibnamefont {Zwiller}}, \bibinfo {author} {\bibfnamefont
  {A.}~\bibnamefont {Rastelli}}, \ and\ \bibinfo {author} {\bibfnamefont
  {R.}~\bibnamefont {Trotta}},\ }\href {\doibase 10.1021/acs.nanolett.7b00777}
  {\bibfield  {journal} {\bibinfo  {journal} {Nano Lett.}\ }\textbf {\bibinfo
  {volume} {17}},\ \bibinfo {pages} {4090} (\bibinfo {year}
  {2017})}\BibitemShut {NoStop}%
\bibitem [{\citenamefont {Ardelt}\ \emph {et~al.}(2014)\citenamefont {Ardelt},
  \citenamefont {Hanschke}, \citenamefont {Fischer}, \citenamefont {M\"uller},
  \citenamefont {Kleinkauf}, \citenamefont {Koller}, \citenamefont {Bechtold},
  \citenamefont {Simmet}, \citenamefont {Wierzbowski}, \citenamefont {Riedl},
  \citenamefont {Abstreiter},\ and\ \citenamefont
  {Finley}}]{Finley_phonon-assisted}%
  \BibitemOpen
  \bibfield  {author} {\bibinfo {author} {\bibfnamefont {P.-L.}\ \bibnamefont
  {Ardelt}}, \bibinfo {author} {\bibfnamefont {L.}~\bibnamefont {Hanschke}},
  \bibinfo {author} {\bibfnamefont {K.~A.}\ \bibnamefont {Fischer}}, \bibinfo
  {author} {\bibfnamefont {K.}~\bibnamefont {M\"uller}}, \bibinfo {author}
  {\bibfnamefont {A.}~\bibnamefont {Kleinkauf}}, \bibinfo {author}
  {\bibfnamefont {M.}~\bibnamefont {Koller}}, \bibinfo {author} {\bibfnamefont
  {A.}~\bibnamefont {Bechtold}}, \bibinfo {author} {\bibfnamefont
  {T.}~\bibnamefont {Simmet}}, \bibinfo {author} {\bibfnamefont
  {J.}~\bibnamefont {Wierzbowski}}, \bibinfo {author} {\bibfnamefont
  {H.}~\bibnamefont {Riedl}}, \bibinfo {author} {\bibfnamefont
  {G.}~\bibnamefont {Abstreiter}}, \ and\ \bibinfo {author} {\bibfnamefont
  {J.~J.}\ \bibnamefont {Finley}},\ }\href {\doibase
  10.1103/PhysRevB.90.241404} {\bibfield  {journal} {\bibinfo  {journal} {Phys.
  Rev. B}\ }\textbf {\bibinfo {volume} {90}},\ \bibinfo {pages} {241404}
  (\bibinfo {year} {2014})}\BibitemShut {NoStop}%
\bibitem [{\citenamefont {Bounouar}\ \emph {et~al.}(2015)\citenamefont
  {Bounouar}, \citenamefont {M\"uller}, \citenamefont {Barth}, \citenamefont
  {Gl\"assl}, \citenamefont {Axt},\ and\ \citenamefont
  {Michler}}]{PI_phonon-assisted_biexc_prep-exp}%
  \BibitemOpen
  \bibfield  {author} {\bibinfo {author} {\bibfnamefont {S.}~\bibnamefont
  {Bounouar}}, \bibinfo {author} {\bibfnamefont {M.}~\bibnamefont {M\"uller}},
  \bibinfo {author} {\bibfnamefont {A.~M.}\ \bibnamefont {Barth}}, \bibinfo
  {author} {\bibfnamefont {M.}~\bibnamefont {Gl\"assl}}, \bibinfo {author}
  {\bibfnamefont {V.~M.}\ \bibnamefont {Axt}}, \ and\ \bibinfo {author}
  {\bibfnamefont {P.}~\bibnamefont {Michler}},\ }\href {\doibase
  10.1103/PhysRevB.91.161302} {\bibfield  {journal} {\bibinfo  {journal} {Phys.
  Rev. B}\ }\textbf {\bibinfo {volume} {91}},\ \bibinfo {pages} {161302(R)}
  (\bibinfo {year} {2015})}\BibitemShut {NoStop}%
\bibitem [{\citenamefont {Gl\"assl}\ \emph
  {et~al.}(2013{\natexlab{a}})\citenamefont {Gl\"assl}, \citenamefont {Barth},\
  and\ \citenamefont {Axt}}]{PI_phonon-assisted_biexc_prep}%
  \BibitemOpen
  \bibfield  {author} {\bibinfo {author} {\bibfnamefont {M.}~\bibnamefont
  {Gl\"assl}}, \bibinfo {author} {\bibfnamefont {A.~M.}\ \bibnamefont {Barth}},
  \ and\ \bibinfo {author} {\bibfnamefont {V.~M.}\ \bibnamefont {Axt}},\ }\href
  {\doibase 10.1103/PhysRevLett.110.147401} {\bibfield  {journal} {\bibinfo
  {journal} {Phys. Rev. Lett.}\ }\textbf {\bibinfo {volume} {110}},\ \bibinfo
  {pages} {147401} (\bibinfo {year} {2013}{\natexlab{a}})}\BibitemShut
  {NoStop}%
\bibitem [{\citenamefont {Barth}\ \emph {et~al.}(2016)\citenamefont {Barth},
  \citenamefont {L\"uker}, \citenamefont {Vagov}, \citenamefont {Reiter},
  \citenamefont {Kuhn},\ and\ \citenamefont {Axt}}]{PI_undressing}%
  \BibitemOpen
  \bibfield  {author} {\bibinfo {author} {\bibfnamefont {A.~M.}\ \bibnamefont
  {Barth}}, \bibinfo {author} {\bibfnamefont {S.}~\bibnamefont {L\"uker}},
  \bibinfo {author} {\bibfnamefont {A.}~\bibnamefont {Vagov}}, \bibinfo
  {author} {\bibfnamefont {D.~E.}\ \bibnamefont {Reiter}}, \bibinfo {author}
  {\bibfnamefont {T.}~\bibnamefont {Kuhn}}, \ and\ \bibinfo {author}
  {\bibfnamefont {V.~M.}\ \bibnamefont {Axt}},\ }\href {\doibase
  10.1103/PhysRevB.94.045306} {\bibfield  {journal} {\bibinfo  {journal} {Phys.
  Rev. B}\ }\textbf {\bibinfo {volume} {94}},\ \bibinfo {pages} {045306}
  (\bibinfo {year} {2016})}\BibitemShut {NoStop}%
\bibitem [{\citenamefont {Reiter}\ \emph {et~al.}(2014)\citenamefont {Reiter},
  \citenamefont {Kuhn}, \citenamefont {Glässl},\ and\ \citenamefont
  {Axt}}]{Reiter_2014}%
  \BibitemOpen
  \bibfield  {author} {\bibinfo {author} {\bibfnamefont {D.~E.}\ \bibnamefont
  {Reiter}}, \bibinfo {author} {\bibfnamefont {T.}~\bibnamefont {Kuhn}},
  \bibinfo {author} {\bibfnamefont {M.}~\bibnamefont {Glässl}}, \ and\
  \bibinfo {author} {\bibfnamefont {V.~M.}\ \bibnamefont {Axt}},\ }\href
  {\doibase 10.1088/0953-8984/26/42/423203} {\bibfield  {journal} {\bibinfo
  {journal} {Journal of Physics: Condensed Matter}\ }\textbf {\bibinfo {volume}
  {26}},\ \bibinfo {pages} {423203} (\bibinfo {year} {2014})}\BibitemShut
  {NoStop}%
\bibitem [{\citenamefont {Debnath}\ \emph {et~al.}(2013)\citenamefont
  {Debnath}, \citenamefont {Meier}, \citenamefont {Chatel},\ and\ \citenamefont
  {Amand}}]{Debnath_ARP_biexciton}%
  \BibitemOpen
  \bibfield  {author} {\bibinfo {author} {\bibfnamefont {A.}~\bibnamefont
  {Debnath}}, \bibinfo {author} {\bibfnamefont {C.}~\bibnamefont {Meier}},
  \bibinfo {author} {\bibfnamefont {B.}~\bibnamefont {Chatel}}, \ and\ \bibinfo
  {author} {\bibfnamefont {T.}~\bibnamefont {Amand}},\ }\href {\doibase
  10.1103/PhysRevB.88.201305} {\bibfield  {journal} {\bibinfo  {journal} {Phys.
  Rev. B}\ }\textbf {\bibinfo {volume} {88}},\ \bibinfo {pages} {201305}
  (\bibinfo {year} {2013})}\BibitemShut {NoStop}%
\bibitem [{\citenamefont {Gl\"assl}\ \emph
  {et~al.}(2013{\natexlab{b}})\citenamefont {Gl\"assl}, \citenamefont {Barth},
  \citenamefont {Gawarecki}, \citenamefont {Machnikowski}, \citenamefont
  {Croitoru}, \citenamefont {L\"uker}, \citenamefont {Reiter}, \citenamefont
  {Kuhn},\ and\ \citenamefont {Axt}}]{Glaessl_ARP_biexciton}%
  \BibitemOpen
  \bibfield  {author} {\bibinfo {author} {\bibfnamefont {M.}~\bibnamefont
  {Gl\"assl}}, \bibinfo {author} {\bibfnamefont {A.~M.}\ \bibnamefont {Barth}},
  \bibinfo {author} {\bibfnamefont {K.}~\bibnamefont {Gawarecki}}, \bibinfo
  {author} {\bibfnamefont {P.}~\bibnamefont {Machnikowski}}, \bibinfo {author}
  {\bibfnamefont {M.~D.}\ \bibnamefont {Croitoru}}, \bibinfo {author}
  {\bibfnamefont {S.}~\bibnamefont {L\"uker}}, \bibinfo {author} {\bibfnamefont
  {D.~E.}\ \bibnamefont {Reiter}}, \bibinfo {author} {\bibfnamefont
  {T.}~\bibnamefont {Kuhn}}, \ and\ \bibinfo {author} {\bibfnamefont {V.~M.}\
  \bibnamefont {Axt}},\ }\href {\doibase 10.1103/PhysRevB.87.085303} {\bibfield
   {journal} {\bibinfo  {journal} {Phys. Rev. B}\ }\textbf {\bibinfo {volume}
  {87}},\ \bibinfo {pages} {085303} (\bibinfo {year}
  {2013}{\natexlab{b}})}\BibitemShut {NoStop}%
\bibitem [{\citenamefont {Kaldewey}\ \emph {et~al.}(2017)\citenamefont
  {Kaldewey}, \citenamefont {L\"uker}, \citenamefont {Kuhlmann}, \citenamefont
  {Valentin}, \citenamefont {Ludwig}, \citenamefont {Wieck}, \citenamefont
  {Reiter}, \citenamefont {Kuhn},\ and\ \citenamefont
  {Warburton}}]{Kaldewey_ARP_biexciton}%
  \BibitemOpen
  \bibfield  {author} {\bibinfo {author} {\bibfnamefont {T.}~\bibnamefont
  {Kaldewey}}, \bibinfo {author} {\bibfnamefont {S.}~\bibnamefont {L\"uker}},
  \bibinfo {author} {\bibfnamefont {A.~V.}\ \bibnamefont {Kuhlmann}}, \bibinfo
  {author} {\bibfnamefont {S.~R.}\ \bibnamefont {Valentin}}, \bibinfo {author}
  {\bibfnamefont {A.}~\bibnamefont {Ludwig}}, \bibinfo {author} {\bibfnamefont
  {A.~D.}\ \bibnamefont {Wieck}}, \bibinfo {author} {\bibfnamefont {D.~E.}\
  \bibnamefont {Reiter}}, \bibinfo {author} {\bibfnamefont {T.}~\bibnamefont
  {Kuhn}}, \ and\ \bibinfo {author} {\bibfnamefont {R.~J.}\ \bibnamefont
  {Warburton}},\ }\href {\doibase 10.1103/PhysRevB.95.161302} {\bibfield
  {journal} {\bibinfo  {journal} {Phys. Rev. B}\ }\textbf {\bibinfo {volume}
  {95}},\ \bibinfo {pages} {161302} (\bibinfo {year} {2017})}\BibitemShut
  {NoStop}%
\bibitem [{\citenamefont {Reiter}\ \emph {et~al.}(2019)\citenamefont {Reiter},
  \citenamefont {Kuhn},\ and\ \citenamefont {Axt}}]{Uebersichtsartikel_2019}%
  \BibitemOpen
  \bibfield  {author} {\bibinfo {author} {\bibfnamefont {D.~E.}\ \bibnamefont
  {Reiter}}, \bibinfo {author} {\bibfnamefont {T.}~\bibnamefont {Kuhn}}, \ and\
  \bibinfo {author} {\bibfnamefont {V.~M.}\ \bibnamefont {Axt}},\ }\href
  {\doibase 10.1080/23746149.2019.1655478} {\bibfield  {journal} {\bibinfo
  {journal} {Advances in Physics: X}\ }\textbf {\bibinfo {volume} {4}},\
  \bibinfo {pages} {1655478} (\bibinfo {year} {2019})}\BibitemShut {NoStop}%
\bibitem [{\citenamefont {Seidelmann}\ \emph
  {et~al.}(2019{\natexlab{a}})\citenamefont {Seidelmann}, \citenamefont
  {Ungar}, \citenamefont {Cygorek}, \citenamefont {Vagov}, \citenamefont
  {Barth}, \citenamefont {Kuhn},\ and\ \citenamefont {Axt}}]{Seidelmann2019}%
  \BibitemOpen
  \bibfield  {author} {\bibinfo {author} {\bibfnamefont {T.}~\bibnamefont
  {Seidelmann}}, \bibinfo {author} {\bibfnamefont {F.}~\bibnamefont {Ungar}},
  \bibinfo {author} {\bibfnamefont {M.}~\bibnamefont {Cygorek}}, \bibinfo
  {author} {\bibfnamefont {A.}~\bibnamefont {Vagov}}, \bibinfo {author}
  {\bibfnamefont {A.~M.}\ \bibnamefont {Barth}}, \bibinfo {author}
  {\bibfnamefont {T.}~\bibnamefont {Kuhn}}, \ and\ \bibinfo {author}
  {\bibfnamefont {V.~M.}\ \bibnamefont {Axt}},\ }\href {\doibase
  10.1103/PhysRevB.99.245301} {\bibfield  {journal} {\bibinfo  {journal} {Phys.
  Rev. B}\ }\textbf {\bibinfo {volume} {99}},\ \bibinfo {pages} {245301}
  (\bibinfo {year} {2019}{\natexlab{a}})}\BibitemShut {NoStop}%
\bibitem [{\citenamefont {Cygorek}\ \emph {et~al.}(2018)\citenamefont
  {Cygorek}, \citenamefont {Ungar}, \citenamefont {Seidelmann}, \citenamefont
  {Barth}, \citenamefont {Vagov}, \citenamefont {Axt},\ and\ \citenamefont
  {Kuhn}}]{Different-Concurrences:18}%
  \BibitemOpen
  \bibfield  {author} {\bibinfo {author} {\bibfnamefont {M.}~\bibnamefont
  {Cygorek}}, \bibinfo {author} {\bibfnamefont {F.}~\bibnamefont {Ungar}},
  \bibinfo {author} {\bibfnamefont {T.}~\bibnamefont {Seidelmann}}, \bibinfo
  {author} {\bibfnamefont {A.~M.}\ \bibnamefont {Barth}}, \bibinfo {author}
  {\bibfnamefont {A.}~\bibnamefont {Vagov}}, \bibinfo {author} {\bibfnamefont
  {V.~M.}\ \bibnamefont {Axt}}, \ and\ \bibinfo {author} {\bibfnamefont
  {T.}~\bibnamefont {Kuhn}},\ }\href {\doibase 10.1103/PhysRevB.98.045303}
  {\bibfield  {journal} {\bibinfo  {journal} {Phys. Rev. B}\ }\textbf {\bibinfo
  {volume} {98}},\ \bibinfo {pages} {045303} (\bibinfo {year}
  {2018})}\BibitemShut {NoStop}%
\bibitem [{\citenamefont {Seidelmann}\ \emph
  {et~al.}(2019{\natexlab{b}})\citenamefont {Seidelmann}, \citenamefont
  {Ungar}, \citenamefont {Barth}, \citenamefont {Vagov}, \citenamefont {Axt},
  \citenamefont {Cygorek},\ and\ \citenamefont
  {Kuhn}}]{Phon_enhanced_entanglement}%
  \BibitemOpen
  \bibfield  {author} {\bibinfo {author} {\bibfnamefont {T.}~\bibnamefont
  {Seidelmann}}, \bibinfo {author} {\bibfnamefont {F.}~\bibnamefont {Ungar}},
  \bibinfo {author} {\bibfnamefont {A.~M.}\ \bibnamefont {Barth}}, \bibinfo
  {author} {\bibfnamefont {A.}~\bibnamefont {Vagov}}, \bibinfo {author}
  {\bibfnamefont {V.~M.}\ \bibnamefont {Axt}}, \bibinfo {author} {\bibfnamefont
  {M.}~\bibnamefont {Cygorek}}, \ and\ \bibinfo {author} {\bibfnamefont
  {T.}~\bibnamefont {Kuhn}},\ }\href {\doibase 10.1103/PhysRevLett.123.137401}
  {\bibfield  {journal} {\bibinfo  {journal} {Phys. Rev. Lett.}\ }\textbf
  {\bibinfo {volume} {123}},\ \bibinfo {pages} {137401} (\bibinfo {year}
  {2019}{\natexlab{b}})}\BibitemShut {NoStop}%
\bibitem [{\citenamefont {Schumacher}\ \emph {et~al.}(2012)\citenamefont
  {Schumacher}, \citenamefont {F\"{o}rstner}, \citenamefont {Zrenner},
  \citenamefont {Florian}, \citenamefont {Gies}, \citenamefont {Gartner},\ and\
  \citenamefont {Jahnke}}]{Jahnke2012}%
  \BibitemOpen
  \bibfield  {author} {\bibinfo {author} {\bibfnamefont {S.}~\bibnamefont
  {Schumacher}}, \bibinfo {author} {\bibfnamefont {J.}~\bibnamefont
  {F\"{o}rstner}}, \bibinfo {author} {\bibfnamefont {A.}~\bibnamefont
  {Zrenner}}, \bibinfo {author} {\bibfnamefont {M.}~\bibnamefont {Florian}},
  \bibinfo {author} {\bibfnamefont {C.}~\bibnamefont {Gies}}, \bibinfo {author}
  {\bibfnamefont {P.}~\bibnamefont {Gartner}}, \ and\ \bibinfo {author}
  {\bibfnamefont {F.}~\bibnamefont {Jahnke}},\ }\href {\doibase
  10.1364/OE.20.005335} {\bibfield  {journal} {\bibinfo  {journal} {Opt.
  Express}\ }\textbf {\bibinfo {volume} {20}},\ \bibinfo {pages} {5335}
  (\bibinfo {year} {2012})}\BibitemShut {NoStop}%
\bibitem [{\citenamefont {Heinze}\ \emph {et~al.}(2017)\citenamefont {Heinze},
  \citenamefont {Zrenner},\ and\ \citenamefont {Schumacher}}]{heinze17}%
  \BibitemOpen
  \bibfield  {author} {\bibinfo {author} {\bibfnamefont {D.}~\bibnamefont
  {Heinze}}, \bibinfo {author} {\bibfnamefont {A.}~\bibnamefont {Zrenner}}, \
  and\ \bibinfo {author} {\bibfnamefont {S.}~\bibnamefont {Schumacher}},\
  }\href {\doibase 10.1103/PhysRevB.95.245306} {\bibfield  {journal} {\bibinfo
  {journal} {Phys. Rev. B}\ }\textbf {\bibinfo {volume} {95}},\ \bibinfo
  {pages} {245306} (\bibinfo {year} {2017})}\BibitemShut {NoStop}%
\bibitem [{\citenamefont {Carmele}\ and\ \citenamefont
  {Knorr}(2011)}]{BiexcCasc_Carmele}%
  \BibitemOpen
  \bibfield  {author} {\bibinfo {author} {\bibfnamefont {A.}~\bibnamefont
  {Carmele}}\ and\ \bibinfo {author} {\bibfnamefont {A.}~\bibnamefont
  {Knorr}},\ }\href {\doibase 10.1103/PhysRevB.84.075328} {\bibfield  {journal}
  {\bibinfo  {journal} {Phys. Rev. B}\ }\textbf {\bibinfo {volume} {84}},\
  \bibinfo {pages} {075328} (\bibinfo {year} {2011})}\BibitemShut {NoStop}%
\bibitem [{\citenamefont {Stevenson}\ \emph {et~al.}(2006)\citenamefont
  {Stevenson}, \citenamefont {Young}, \citenamefont {Atkinson}, \citenamefont
  {Cooper}, \citenamefont {Ritchie},\ and\ \citenamefont
  {Shields}}]{Stevenson2006}%
  \BibitemOpen
  \bibfield  {author} {\bibinfo {author} {\bibfnamefont {R.~M.}\ \bibnamefont
  {Stevenson}}, \bibinfo {author} {\bibfnamefont {R.~J.}\ \bibnamefont
  {Young}}, \bibinfo {author} {\bibfnamefont {P.}~\bibnamefont {Atkinson}},
  \bibinfo {author} {\bibfnamefont {K.}~\bibnamefont {Cooper}}, \bibinfo
  {author} {\bibfnamefont {D.~A.}\ \bibnamefont {Ritchie}}, \ and\ \bibinfo
  {author} {\bibfnamefont {A.~J.}\ \bibnamefont {Shields}},\ }\href
  {http://dx.doi.org/10.1038/nature04446} {\bibfield  {journal} {\bibinfo
  {journal} {Nature}\ }\textbf {\bibinfo {volume} {439}},\ \bibinfo {pages}
  {179} (\bibinfo {year} {2006})}\BibitemShut {NoStop}%
\bibitem [{\citenamefont {Young}\ \emph {et~al.}(2006)\citenamefont {Young},
  \citenamefont {Stevenson}, \citenamefont {Atkinson}, \citenamefont {Cooper},
  \citenamefont {Ritchie},\ and\ \citenamefont {Shields}}]{Young_2006}%
  \BibitemOpen
  \bibfield  {author} {\bibinfo {author} {\bibfnamefont {R.~J.}\ \bibnamefont
  {Young}}, \bibinfo {author} {\bibfnamefont {R.~M.}\ \bibnamefont
  {Stevenson}}, \bibinfo {author} {\bibfnamefont {P.}~\bibnamefont {Atkinson}},
  \bibinfo {author} {\bibfnamefont {K.}~\bibnamefont {Cooper}}, \bibinfo
  {author} {\bibfnamefont {D.~A.}\ \bibnamefont {Ritchie}}, \ and\ \bibinfo
  {author} {\bibfnamefont {A.~J.}\ \bibnamefont {Shields}},\ }\href {\doibase
  10.1088/1367-2630/8/2/029} {\bibfield  {journal} {\bibinfo  {journal} {New
  Journal of Physics}\ }\textbf {\bibinfo {volume} {8}},\ \bibinfo {pages} {29}
  (\bibinfo {year} {2006})}\BibitemShut {NoStop}%
\bibitem [{\citenamefont {Muller}\ \emph {et~al.}(2009)\citenamefont {Muller},
  \citenamefont {Fang}, \citenamefont {Lawall},\ and\ \citenamefont
  {Solomon}}]{Muller_2009}%
  \BibitemOpen
  \bibfield  {author} {\bibinfo {author} {\bibfnamefont {A.}~\bibnamefont
  {Muller}}, \bibinfo {author} {\bibfnamefont {W.}~\bibnamefont {Fang}},
  \bibinfo {author} {\bibfnamefont {J.}~\bibnamefont {Lawall}}, \ and\ \bibinfo
  {author} {\bibfnamefont {G.~S.}\ \bibnamefont {Solomon}},\ }\href {\doibase
  10.1103/PhysRevLett.103.217402} {\bibfield  {journal} {\bibinfo  {journal}
  {Phys. Rev. Lett.}\ }\textbf {\bibinfo {volume} {103}},\ \bibinfo {pages}
  {217402} (\bibinfo {year} {2009})}\BibitemShut {NoStop}%
\bibitem [{\citenamefont {Huber}\ \emph
  {et~al.}(2018{\natexlab{b}})\citenamefont {Huber}, \citenamefont {Reindl},
  \citenamefont {Covre~da Silva}, \citenamefont {Schimpf}, \citenamefont
  {Mart\'{\i}n-S\'anchez}, \citenamefont {Huang}, \citenamefont {Piredda},
  \citenamefont {Edlinger}, \citenamefont {Rastelli},\ and\ \citenamefont
  {Trotta}}]{Huber_PRL_2018}%
  \BibitemOpen
  \bibfield  {author} {\bibinfo {author} {\bibfnamefont {D.}~\bibnamefont
  {Huber}}, \bibinfo {author} {\bibfnamefont {M.}~\bibnamefont {Reindl}},
  \bibinfo {author} {\bibfnamefont {S.~F.}\ \bibnamefont {Covre~da Silva}},
  \bibinfo {author} {\bibfnamefont {C.}~\bibnamefont {Schimpf}}, \bibinfo
  {author} {\bibfnamefont {J.}~\bibnamefont {Mart\'{\i}n-S\'anchez}}, \bibinfo
  {author} {\bibfnamefont {H.}~\bibnamefont {Huang}}, \bibinfo {author}
  {\bibfnamefont {G.}~\bibnamefont {Piredda}}, \bibinfo {author} {\bibfnamefont
  {J.}~\bibnamefont {Edlinger}}, \bibinfo {author} {\bibfnamefont
  {A.}~\bibnamefont {Rastelli}}, \ and\ \bibinfo {author} {\bibfnamefont
  {R.}~\bibnamefont {Trotta}},\ }\href {\doibase
  10.1103/PhysRevLett.121.033902} {\bibfield  {journal} {\bibinfo  {journal}
  {Phys. Rev. Lett.}\ }\textbf {\bibinfo {volume} {121}},\ \bibinfo {pages}
  {033902} (\bibinfo {year} {2018}{\natexlab{b}})}\BibitemShut {NoStop}%
\bibitem [{\citenamefont {Wang}\ \emph {et~al.}(2019)\citenamefont {Wang},
  \citenamefont {Hu}, \citenamefont {Chung}, \citenamefont {Qin}, \citenamefont
  {Yang}, \citenamefont {Li}, \citenamefont {Liu}, \citenamefont {Zhong},
  \citenamefont {He}, \citenamefont {Ding}, \citenamefont {Deng}, \citenamefont
  {Dai}, \citenamefont {Huo}, \citenamefont {H\"ofling}, \citenamefont {Lu},\
  and\ \citenamefont {Pan}}]{Wang_2019}%
  \BibitemOpen
  \bibfield  {author} {\bibinfo {author} {\bibfnamefont {H.}~\bibnamefont
  {Wang}}, \bibinfo {author} {\bibfnamefont {H.}~\bibnamefont {Hu}}, \bibinfo
  {author} {\bibfnamefont {T.-H.}\ \bibnamefont {Chung}}, \bibinfo {author}
  {\bibfnamefont {J.}~\bibnamefont {Qin}}, \bibinfo {author} {\bibfnamefont
  {X.}~\bibnamefont {Yang}}, \bibinfo {author} {\bibfnamefont {J.-P.}\
  \bibnamefont {Li}}, \bibinfo {author} {\bibfnamefont {R.-Z.}\ \bibnamefont
  {Liu}}, \bibinfo {author} {\bibfnamefont {H.-S.}\ \bibnamefont {Zhong}},
  \bibinfo {author} {\bibfnamefont {Y.-M.}\ \bibnamefont {He}}, \bibinfo
  {author} {\bibfnamefont {X.}~\bibnamefont {Ding}}, \bibinfo {author}
  {\bibfnamefont {Y.-H.}\ \bibnamefont {Deng}}, \bibinfo {author}
  {\bibfnamefont {Q.}~\bibnamefont {Dai}}, \bibinfo {author} {\bibfnamefont
  {Y.-H.}\ \bibnamefont {Huo}}, \bibinfo {author} {\bibfnamefont
  {S.}~\bibnamefont {H\"ofling}}, \bibinfo {author} {\bibfnamefont {C.-Y.}\
  \bibnamefont {Lu}}, \ and\ \bibinfo {author} {\bibfnamefont {J.-W.}\
  \bibnamefont {Pan}},\ }\href {\doibase 10.1103/PhysRevLett.122.113602}
  {\bibfield  {journal} {\bibinfo  {journal} {Phys. Rev. Lett.}\ }\textbf
  {\bibinfo {volume} {122}},\ \bibinfo {pages} {113602} (\bibinfo {year}
  {2019})}\BibitemShut {NoStop}%
\bibitem [{\citenamefont {Liu}\ \emph {et~al.}(2019)\citenamefont {Liu},
  \citenamefont {Su}, \citenamefont {Wei}, \citenamefont {Yao}, \citenamefont
  {Silva}, \citenamefont {Yu}, \citenamefont {Iles-Smith}, \citenamefont
  {Srinivasan}, \citenamefont {Rastelli}, \citenamefont {Li},\ and\
  \citenamefont {Wang}}]{Liu2019}%
  \BibitemOpen
  \bibfield  {author} {\bibinfo {author} {\bibfnamefont {J.}~\bibnamefont
  {Liu}}, \bibinfo {author} {\bibfnamefont {R.}~\bibnamefont {Su}}, \bibinfo
  {author} {\bibfnamefont {Y.}~\bibnamefont {Wei}}, \bibinfo {author}
  {\bibfnamefont {B.}~\bibnamefont {Yao}}, \bibinfo {author} {\bibfnamefont
  {S.~F. C.~d.}\ \bibnamefont {Silva}}, \bibinfo {author} {\bibfnamefont
  {Y.}~\bibnamefont {Yu}}, \bibinfo {author} {\bibfnamefont {J.}~\bibnamefont
  {Iles-Smith}}, \bibinfo {author} {\bibfnamefont {K.}~\bibnamefont
  {Srinivasan}}, \bibinfo {author} {\bibfnamefont {A.}~\bibnamefont
  {Rastelli}}, \bibinfo {author} {\bibfnamefont {J.}~\bibnamefont {Li}}, \ and\
  \bibinfo {author} {\bibfnamefont {X.}~\bibnamefont {Wang}},\ }\href {\doibase
  10.1038/s41565-019-0435-9} {\bibfield  {journal} {\bibinfo  {journal} {Nature
  Nanotechnology}\ }\textbf {\bibinfo {volume} {14}},\ \bibinfo {pages} {586}
  (\bibinfo {year} {2019})}\BibitemShut {NoStop}%
\bibitem [{\citenamefont {Bounouar}\ \emph {et~al.}(2018)\citenamefont
  {Bounouar}, \citenamefont {de~la Haye}, \citenamefont {Strauß},
  \citenamefont {Schnauber}, \citenamefont {Thoma}, \citenamefont {Gschrey},
  \citenamefont {Schulze}, \citenamefont {Strittmatter}, \citenamefont {Rodt},\
  and\ \citenamefont {Reitzenstein}}]{Bounouar18}%
  \BibitemOpen
  \bibfield  {author} {\bibinfo {author} {\bibfnamefont {S.}~\bibnamefont
  {Bounouar}}, \bibinfo {author} {\bibfnamefont {C.}~\bibnamefont {de~la
  Haye}}, \bibinfo {author} {\bibfnamefont {M.}~\bibnamefont {Strauß}},
  \bibinfo {author} {\bibfnamefont {P.}~\bibnamefont {Schnauber}}, \bibinfo
  {author} {\bibfnamefont {A.}~\bibnamefont {Thoma}}, \bibinfo {author}
  {\bibfnamefont {M.}~\bibnamefont {Gschrey}}, \bibinfo {author} {\bibfnamefont
  {J.-H.}\ \bibnamefont {Schulze}}, \bibinfo {author} {\bibfnamefont
  {A.}~\bibnamefont {Strittmatter}}, \bibinfo {author} {\bibfnamefont
  {S.}~\bibnamefont {Rodt}}, \ and\ \bibinfo {author} {\bibfnamefont
  {S.}~\bibnamefont {Reitzenstein}},\ }\href {\doibase 10.1063/1.5020242}
  {\bibfield  {journal} {\bibinfo  {journal} {Appl. Phys. Lett.}\ }\textbf
  {\bibinfo {volume} {112}},\ \bibinfo {pages} {153107} (\bibinfo {year}
  {2018})}\BibitemShut {NoStop}%
\bibitem [{\citenamefont {Dousse}\ \emph {et~al.}(2010)\citenamefont {Dousse},
  \citenamefont {Suffczy{\'n}ski}, \citenamefont {Beveratos}, \citenamefont
  {Krebs}, \citenamefont {Lema{\^i}tre}, \citenamefont {Sagnes}, \citenamefont
  {Bloch}, \citenamefont {Voisin},\ and\ \citenamefont
  {Senellart}}]{dousse:10}%
  \BibitemOpen
  \bibfield  {author} {\bibinfo {author} {\bibfnamefont {A.}~\bibnamefont
  {Dousse}}, \bibinfo {author} {\bibfnamefont {J.}~\bibnamefont
  {Suffczy{\'n}ski}}, \bibinfo {author} {\bibfnamefont {A.}~\bibnamefont
  {Beveratos}}, \bibinfo {author} {\bibfnamefont {O.}~\bibnamefont {Krebs}},
  \bibinfo {author} {\bibfnamefont {A.}~\bibnamefont {Lema{\^i}tre}}, \bibinfo
  {author} {\bibfnamefont {I.}~\bibnamefont {Sagnes}}, \bibinfo {author}
  {\bibfnamefont {J.}~\bibnamefont {Bloch}}, \bibinfo {author} {\bibfnamefont
  {P.}~\bibnamefont {Voisin}}, \ and\ \bibinfo {author} {\bibfnamefont
  {P.}~\bibnamefont {Senellart}},\ }\href {https://doi.org/10.1038/nature09148}
  {\bibfield  {journal} {\bibinfo  {journal} {Nature}\ }\textbf {\bibinfo
  {volume} {466}},\ \bibinfo {pages} {217} (\bibinfo {year}
  {2010})}\BibitemShut {NoStop}%
\bibitem [{\citenamefont {Winik}\ \emph {et~al.}(2017)\citenamefont {Winik},
  \citenamefont {Cogan}, \citenamefont {Don}, \citenamefont {Schwartz},
  \citenamefont {Gantz}, \citenamefont {Schmidgall}, \citenamefont {Livneh},
  \citenamefont {Rapaport}, \citenamefont {Buks},\ and\ \citenamefont
  {Gershoni}}]{winik:2017}%
  \BibitemOpen
  \bibfield  {author} {\bibinfo {author} {\bibfnamefont {R.}~\bibnamefont
  {Winik}}, \bibinfo {author} {\bibfnamefont {D.}~\bibnamefont {Cogan}},
  \bibinfo {author} {\bibfnamefont {Y.}~\bibnamefont {Don}}, \bibinfo {author}
  {\bibfnamefont {I.}~\bibnamefont {Schwartz}}, \bibinfo {author}
  {\bibfnamefont {L.}~\bibnamefont {Gantz}}, \bibinfo {author} {\bibfnamefont
  {E.~R.}\ \bibnamefont {Schmidgall}}, \bibinfo {author} {\bibfnamefont
  {N.}~\bibnamefont {Livneh}}, \bibinfo {author} {\bibfnamefont
  {R.}~\bibnamefont {Rapaport}}, \bibinfo {author} {\bibfnamefont
  {E.}~\bibnamefont {Buks}}, \ and\ \bibinfo {author} {\bibfnamefont
  {D.}~\bibnamefont {Gershoni}},\ }\href {\doibase 10.1103/PhysRevB.95.235435}
  {\bibfield  {journal} {\bibinfo  {journal} {Phys. Rev. B}\ }\textbf {\bibinfo
  {volume} {95}},\ \bibinfo {pages} {235435} (\bibinfo {year}
  {2017})}\BibitemShut {NoStop}%
\bibitem [{\citenamefont {Fognini}\ \emph {et~al.}(2019)\citenamefont
  {Fognini}, \citenamefont {Ahmadi}, \citenamefont {Zeeshan}, \citenamefont
  {Fokkens}, \citenamefont {Gibson}, \citenamefont {Sherlekar}, \citenamefont
  {Daley}, \citenamefont {Dalacu}, \citenamefont {Poole}, \citenamefont
  {J\"ons}, \citenamefont {Zwiller},\ and\ \citenamefont
  {Reimer}}]{Fognini_2019}%
  \BibitemOpen
  \bibfield  {author} {\bibinfo {author} {\bibfnamefont {A.}~\bibnamefont
  {Fognini}}, \bibinfo {author} {\bibfnamefont {A.}~\bibnamefont {Ahmadi}},
  \bibinfo {author} {\bibfnamefont {M.}~\bibnamefont {Zeeshan}}, \bibinfo
  {author} {\bibfnamefont {J.~T.}\ \bibnamefont {Fokkens}}, \bibinfo {author}
  {\bibfnamefont {S.~J.}\ \bibnamefont {Gibson}}, \bibinfo {author}
  {\bibfnamefont {N.}~\bibnamefont {Sherlekar}}, \bibinfo {author}
  {\bibfnamefont {S.~J.}\ \bibnamefont {Daley}}, \bibinfo {author}
  {\bibfnamefont {D.}~\bibnamefont {Dalacu}}, \bibinfo {author} {\bibfnamefont
  {P.~J.}\ \bibnamefont {Poole}}, \bibinfo {author} {\bibfnamefont {K.~D.}\
  \bibnamefont {J\"ons}}, \bibinfo {author} {\bibfnamefont {V.}~\bibnamefont
  {Zwiller}}, \ and\ \bibinfo {author} {\bibfnamefont {M.~E.}\ \bibnamefont
  {Reimer}},\ }\href {\doibase 10.1021/acsphotonics.8b01496} {\bibfield
  {journal} {\bibinfo  {journal} {ACS Photonics}\ }\textbf {\bibinfo {volume}
  {6}},\ \bibinfo {pages} {1656} (\bibinfo {year} {2019})}\BibitemShut
  {NoStop}%
\bibitem [{\citenamefont {Hafenbrak}\ \emph {et~al.}(2007)\citenamefont
  {Hafenbrak}, \citenamefont {Ulrich}, \citenamefont {Michler}, \citenamefont
  {Wang}, \citenamefont {Rastelli},\ and\ \citenamefont {Schmidt}}]{Hafenbrak}%
  \BibitemOpen
  \bibfield  {author} {\bibinfo {author} {\bibfnamefont {R.}~\bibnamefont
  {Hafenbrak}}, \bibinfo {author} {\bibfnamefont {S.~M.}\ \bibnamefont
  {Ulrich}}, \bibinfo {author} {\bibfnamefont {P.}~\bibnamefont {Michler}},
  \bibinfo {author} {\bibfnamefont {L.}~\bibnamefont {Wang}}, \bibinfo {author}
  {\bibfnamefont {A.}~\bibnamefont {Rastelli}}, \ and\ \bibinfo {author}
  {\bibfnamefont {O.~G.}\ \bibnamefont {Schmidt}},\ }\href
  {http://stacks.iop.org/1367-2630/9/i=9/a=315} {\bibfield  {journal} {\bibinfo
   {journal} {New Journal of Physics}\ }\textbf {\bibinfo {volume} {9}},\
  \bibinfo {pages} {315} (\bibinfo {year} {2007})}\BibitemShut {NoStop}%
\bibitem [{\citenamefont {Bennett}\ \emph {et~al.}(2010)\citenamefont
  {Bennett}, \citenamefont {Pooley}, \citenamefont {Stevenson}, \citenamefont
  {Ward}, \citenamefont {Patel}, \citenamefont {{Boyer de la Giroday}},
  \citenamefont {Sk{\"o}ld}, \citenamefont {Farrer}, \citenamefont {Nicoll},
  \citenamefont {Ritchie},\ and\ \citenamefont
  {Shields}}]{Biexc_FSS_electrical_control_Bennett}%
  \BibitemOpen
  \bibfield  {author} {\bibinfo {author} {\bibfnamefont {A.~J.}\ \bibnamefont
  {Bennett}}, \bibinfo {author} {\bibfnamefont {M.~A.}\ \bibnamefont {Pooley}},
  \bibinfo {author} {\bibfnamefont {R.~M.}\ \bibnamefont {Stevenson}}, \bibinfo
  {author} {\bibfnamefont {M.~B.}\ \bibnamefont {Ward}}, \bibinfo {author}
  {\bibfnamefont {R.~B.}\ \bibnamefont {Patel}}, \bibinfo {author}
  {\bibfnamefont {A.}~\bibnamefont {{Boyer de la Giroday}}}, \bibinfo {author}
  {\bibfnamefont {N.}~\bibnamefont {Sk{\"o}ld}}, \bibinfo {author}
  {\bibfnamefont {I.}~\bibnamefont {Farrer}}, \bibinfo {author} {\bibfnamefont
  {C.~A.}\ \bibnamefont {Nicoll}}, \bibinfo {author} {\bibfnamefont {D.~A.}\
  \bibnamefont {Ritchie}}, \ and\ \bibinfo {author} {\bibfnamefont {A.~J.}\
  \bibnamefont {Shields}},\ }\href {\doibase 10.1038/nphys1780} {\bibfield
  {journal} {\bibinfo  {journal} {Nature Physics}\ }\textbf {\bibinfo {volume}
  {6}},\ \bibinfo {pages} {947} (\bibinfo {year} {2010})}\BibitemShut {NoStop}%
\bibitem [{\citenamefont {del Valle}(2013)}]{EdV}%
  \BibitemOpen
  \bibfield  {author} {\bibinfo {author} {\bibfnamefont {E.}~\bibnamefont {del
  Valle}},\ }\href {http://stacks.iop.org/1367-2630/15/i=2/a=025019} {\bibfield
   {journal} {\bibinfo  {journal} {New J. Phys.}\ }\textbf {\bibinfo {volume}
  {15}},\ \bibinfo {pages} {025019} (\bibinfo {year} {2013})}\BibitemShut
  {NoStop}%
\bibitem [{\citenamefont {Troiani}\ \emph {et~al.}(2006)\citenamefont
  {Troiani}, \citenamefont {Perea},\ and\ \citenamefont
  {Tejedor}}]{Troiani2006}%
  \BibitemOpen
  \bibfield  {author} {\bibinfo {author} {\bibfnamefont {F.}~\bibnamefont
  {Troiani}}, \bibinfo {author} {\bibfnamefont {J.~I.}\ \bibnamefont {Perea}},
  \ and\ \bibinfo {author} {\bibfnamefont {C.}~\bibnamefont {Tejedor}},\ }\href
  {\doibase 10.1103/PhysRevB.74.235310} {\bibfield  {journal} {\bibinfo
  {journal} {Phys. Rev. B}\ }\textbf {\bibinfo {volume} {74}},\ \bibinfo
  {pages} {235310} (\bibinfo {year} {2006})}\BibitemShut {NoStop}%
\bibitem [{\citenamefont {Stevenson}\ \emph {et~al.}(2012)\citenamefont
  {Stevenson}, \citenamefont {Salter}, \citenamefont {Nilsson}, \citenamefont
  {Bennett}, \citenamefont {Ward}, \citenamefont {Farrer}, \citenamefont
  {Ritchie},\ and\ \citenamefont {Shields}}]{stevenson:2012}%
  \BibitemOpen
  \bibfield  {author} {\bibinfo {author} {\bibfnamefont {R.~M.}\ \bibnamefont
  {Stevenson}}, \bibinfo {author} {\bibfnamefont {C.~L.}\ \bibnamefont
  {Salter}}, \bibinfo {author} {\bibfnamefont {J.}~\bibnamefont {Nilsson}},
  \bibinfo {author} {\bibfnamefont {A.~J.}\ \bibnamefont {Bennett}}, \bibinfo
  {author} {\bibfnamefont {M.~B.}\ \bibnamefont {Ward}}, \bibinfo {author}
  {\bibfnamefont {I.}~\bibnamefont {Farrer}}, \bibinfo {author} {\bibfnamefont
  {D.~A.}\ \bibnamefont {Ritchie}}, \ and\ \bibinfo {author} {\bibfnamefont
  {A.~J.}\ \bibnamefont {Shields}},\ }\href {\doibase
  10.1103/PhysRevLett.108.040503} {\bibfield  {journal} {\bibinfo  {journal}
  {Phys. Rev. Lett.}\ }\textbf {\bibinfo {volume} {108}},\ \bibinfo {pages}
  {040503} (\bibinfo {year} {2012})}\BibitemShut {NoStop}%
\bibitem [{\citenamefont {Benson}\ \emph {et~al.}(2000)\citenamefont {Benson},
  \citenamefont {Santori}, \citenamefont {Pelton},\ and\ \citenamefont
  {Yamamoto}}]{Benson_2000_QD_cav_device}%
  \BibitemOpen
  \bibfield  {author} {\bibinfo {author} {\bibfnamefont {O.}~\bibnamefont
  {Benson}}, \bibinfo {author} {\bibfnamefont {C.}~\bibnamefont {Santori}},
  \bibinfo {author} {\bibfnamefont {M.}~\bibnamefont {Pelton}}, \ and\ \bibinfo
  {author} {\bibfnamefont {Y.}~\bibnamefont {Yamamoto}},\ }\href {\doibase
  10.1103/PhysRevLett.84.2513} {\bibfield  {journal} {\bibinfo  {journal}
  {Phys. Rev. Lett.}\ }\textbf {\bibinfo {volume} {84}},\ \bibinfo {pages}
  {2513} (\bibinfo {year} {2000})}\BibitemShut {NoStop}%
\bibitem [{\citenamefont {S{\'a}nchez~Mu{\~n}oz}\ \emph
  {et~al.}(2015)\citenamefont {S{\'a}nchez~Mu{\~n}oz}, \citenamefont {Laussy},
  \citenamefont {Tejedor},\ and\ \citenamefont {del Valle}}]{munoz15}%
  \BibitemOpen
  \bibfield  {author} {\bibinfo {author} {\bibfnamefont {C.}~\bibnamefont
  {S{\'a}nchez~Mu{\~n}oz}}, \bibinfo {author} {\bibfnamefont {F.~P.}\
  \bibnamefont {Laussy}}, \bibinfo {author} {\bibfnamefont {C.}~\bibnamefont
  {Tejedor}}, \ and\ \bibinfo {author} {\bibfnamefont {E.}~\bibnamefont {del
  Valle}},\ }\href {http://stacks.iop.org/1367-2630/17/i=12/a=123021}
  {\bibfield  {journal} {\bibinfo  {journal} {New J. Phys.}\ }\textbf {\bibinfo
  {volume} {17}},\ \bibinfo {pages} {123021} (\bibinfo {year}
  {2015})}\BibitemShut {NoStop}%
\bibitem [{\citenamefont {Badolato}\ \emph {et~al.}(2005)\citenamefont
  {Badolato}, \citenamefont {Hennessy}, \citenamefont {Atat{\"u}re},
  \citenamefont {Dreiser}, \citenamefont {Hu}, \citenamefont {Petroff},\ and\
  \citenamefont {Imamo{\u g}lu}}]{Badolato:05}%
  \BibitemOpen
  \bibfield  {author} {\bibinfo {author} {\bibfnamefont {A.}~\bibnamefont
  {Badolato}}, \bibinfo {author} {\bibfnamefont {K.}~\bibnamefont {Hennessy}},
  \bibinfo {author} {\bibfnamefont {M.}~\bibnamefont {Atat{\"u}re}}, \bibinfo
  {author} {\bibfnamefont {J.}~\bibnamefont {Dreiser}}, \bibinfo {author}
  {\bibfnamefont {E.}~\bibnamefont {Hu}}, \bibinfo {author} {\bibfnamefont
  {P.~M.}\ \bibnamefont {Petroff}}, \ and\ \bibinfo {author} {\bibfnamefont
  {A.}~\bibnamefont {Imamo{\u g}lu}},\ }\href {\doibase
  10.1126/science.1109815} {\bibfield  {journal} {\bibinfo  {journal}
  {Science}\ }\textbf {\bibinfo {volume} {308}},\ \bibinfo {pages} {1158}
  (\bibinfo {year} {2005})}\BibitemShut {NoStop}%
\bibitem [{\citenamefont {Ota}\ \emph {et~al.}(2011)\citenamefont {Ota},
  \citenamefont {Iwamoto}, \citenamefont {Kumagai},\ and\ \citenamefont
  {Arakawa}}]{Ota_2pht_emis_cavity_2011}%
  \BibitemOpen
  \bibfield  {author} {\bibinfo {author} {\bibfnamefont {Y.}~\bibnamefont
  {Ota}}, \bibinfo {author} {\bibfnamefont {S.}~\bibnamefont {Iwamoto}},
  \bibinfo {author} {\bibfnamefont {N.}~\bibnamefont {Kumagai}}, \ and\
  \bibinfo {author} {\bibfnamefont {Y.}~\bibnamefont {Arakawa}},\ }\href
  {\doibase 10.1103/PhysRevLett.107.233602} {\bibfield  {journal} {\bibinfo
  {journal} {Phys. Rev. Lett.}\ }\textbf {\bibinfo {volume} {107}},\ \bibinfo
  {pages} {233602} (\bibinfo {year} {2011})}\BibitemShut {NoStop}%
\bibitem [{\citenamefont {Mermillod}\ \emph {et~al.}(2016)\citenamefont
  {Mermillod}, \citenamefont {Wigger}, \citenamefont {Delmonte}, \citenamefont
  {Reiter}, \citenamefont {Schneider}, \citenamefont {Kamp}, \citenamefont
  {H\"{o}fling}, \citenamefont {Langbein}, \citenamefont {Kuhn}, \citenamefont
  {Nogues},\ and\ \citenamefont {Kasprzak}}]{Mermillod:16}%
  \BibitemOpen
  \bibfield  {author} {\bibinfo {author} {\bibfnamefont {Q.}~\bibnamefont
  {Mermillod}}, \bibinfo {author} {\bibfnamefont {D.}~\bibnamefont {Wigger}},
  \bibinfo {author} {\bibfnamefont {V.}~\bibnamefont {Delmonte}}, \bibinfo
  {author} {\bibfnamefont {D.~E.}\ \bibnamefont {Reiter}}, \bibinfo {author}
  {\bibfnamefont {C.}~\bibnamefont {Schneider}}, \bibinfo {author}
  {\bibfnamefont {M.}~\bibnamefont {Kamp}}, \bibinfo {author} {\bibfnamefont
  {S.}~\bibnamefont {H\"{o}fling}}, \bibinfo {author} {\bibfnamefont
  {W.}~\bibnamefont {Langbein}}, \bibinfo {author} {\bibfnamefont
  {T.}~\bibnamefont {Kuhn}}, \bibinfo {author} {\bibfnamefont {G.}~\bibnamefont
  {Nogues}}, \ and\ \bibinfo {author} {\bibfnamefont {J.}~\bibnamefont
  {Kasprzak}},\ }\href {\doibase 10.1364/OPTICA.3.000377} {\bibfield  {journal}
  {\bibinfo  {journal} {Optica}\ }\textbf {\bibinfo {volume} {3}},\ \bibinfo
  {pages} {377} (\bibinfo {year} {2016})}\BibitemShut {NoStop}%
\bibitem [{\citenamefont {James}\ \emph {et~al.}(2001)\citenamefont {James},
  \citenamefont {Kwiat}, \citenamefont {Munro},\ and\ \citenamefont
  {White}}]{QuantumStateTomography}%
  \BibitemOpen
  \bibfield  {author} {\bibinfo {author} {\bibfnamefont {D.~F.~V.}\
  \bibnamefont {James}}, \bibinfo {author} {\bibfnamefont {P.~G.}\ \bibnamefont
  {Kwiat}}, \bibinfo {author} {\bibfnamefont {W.~J.}\ \bibnamefont {Munro}}, \
  and\ \bibinfo {author} {\bibfnamefont {A.~G.}\ \bibnamefont {White}},\ }\href
  {\doibase 10.1103/PhysRevA.64.052312} {\bibfield  {journal} {\bibinfo
  {journal} {Phys. Rev. A}\ }\textbf {\bibinfo {volume} {64}},\ \bibinfo
  {pages} {052312} (\bibinfo {year} {2001})}\BibitemShut {NoStop}%
\bibitem [{\citenamefont {Stevenson}\ \emph {et~al.}(2008)\citenamefont
  {Stevenson}, \citenamefont {Hudson}, \citenamefont {Bennett}, \citenamefont
  {Young}, \citenamefont {Nicoll}, \citenamefont {Ritchie},\ and\ \citenamefont
  {Shields}}]{StevensonPRL2008}%
  \BibitemOpen
  \bibfield  {author} {\bibinfo {author} {\bibfnamefont {R.~M.}\ \bibnamefont
  {Stevenson}}, \bibinfo {author} {\bibfnamefont {A.~J.}\ \bibnamefont
  {Hudson}}, \bibinfo {author} {\bibfnamefont {A.~J.}\ \bibnamefont {Bennett}},
  \bibinfo {author} {\bibfnamefont {R.~J.}\ \bibnamefont {Young}}, \bibinfo
  {author} {\bibfnamefont {C.~A.}\ \bibnamefont {Nicoll}}, \bibinfo {author}
  {\bibfnamefont {D.~A.}\ \bibnamefont {Ritchie}}, \ and\ \bibinfo {author}
  {\bibfnamefont {A.~J.}\ \bibnamefont {Shields}},\ }\href {\doibase
  10.1103/PhysRevLett.101.170501} {\bibfield  {journal} {\bibinfo  {journal}
  {Phys. Rev. Lett.}\ }\textbf {\bibinfo {volume} {101}},\ \bibinfo {pages}
  {170501} (\bibinfo {year} {2008})}\BibitemShut {NoStop}%
\bibitem [{\citenamefont {Cosacchi}\ \emph {et~al.}(2018)\citenamefont
  {Cosacchi}, \citenamefont {Cygorek}, \citenamefont {Ungar}, \citenamefont
  {Barth}, \citenamefont {Vagov},\ and\ \citenamefont {Axt}}]{multi-time}%
  \BibitemOpen
  \bibfield  {author} {\bibinfo {author} {\bibfnamefont {M.}~\bibnamefont
  {Cosacchi}}, \bibinfo {author} {\bibfnamefont {M.}~\bibnamefont {Cygorek}},
  \bibinfo {author} {\bibfnamefont {F.}~\bibnamefont {Ungar}}, \bibinfo
  {author} {\bibfnamefont {A.~M.}\ \bibnamefont {Barth}}, \bibinfo {author}
  {\bibfnamefont {A.}~\bibnamefont {Vagov}}, \ and\ \bibinfo {author}
  {\bibfnamefont {V.~M.}\ \bibnamefont {Axt}},\ }\href
  {https://link.aps.org/doi/10.1103/PhysRevB.98.125302} {\bibfield  {journal}
  {\bibinfo  {journal} {Phys. Rev. B}\ }\textbf {\bibinfo {volume} {98}},\
  \bibinfo {pages} {125302} (\bibinfo {year} {2018})}\BibitemShut {NoStop}%
\bibitem [{\citenamefont {Wootters}(1998)}]{Wootters1998}%
  \BibitemOpen
  \bibfield  {author} {\bibinfo {author} {\bibfnamefont {W.~K.}\ \bibnamefont
  {Wootters}},\ }\href {\doibase 10.1103/PhysRevLett.80.2245} {\bibfield
  {journal} {\bibinfo  {journal} {Phys. Rev. Lett.}\ }\textbf {\bibinfo
  {volume} {80}},\ \bibinfo {pages} {2245} (\bibinfo {year}
  {1998})}\BibitemShut {NoStop}%
\bibitem [{\citenamefont {Winkler}(2003)}]{Winkler}%
  \BibitemOpen
  \bibfield  {author} {\bibinfo {author} {\bibfnamefont {R.}~\bibnamefont
  {Winkler}},\ }\href@noop {} {\emph {\bibinfo {title} {Spin-Orbit Coupling
  Effects in Two-Dimensional Electron and Hole Systems}}},\ \bibinfo {series}
  {Springer Tracts in Modern Physics}, Vol.\ \bibinfo {volume} {191}\ (\bibinfo
   {publisher} {Springer},\ \bibinfo {year} {2003})\BibitemShut {NoStop}%
\bibitem [{\citenamefont {Bravyi}\ \emph {et~al.}(2011)\citenamefont {Bravyi},
  \citenamefont {DiVincenzo},\ and\ \citenamefont {Loss}}]{LossDivincenzo_SW}%
  \BibitemOpen
  \bibfield  {author} {\bibinfo {author} {\bibfnamefont {S.}~\bibnamefont
  {Bravyi}}, \bibinfo {author} {\bibfnamefont {D.~P.}\ \bibnamefont
  {DiVincenzo}}, \ and\ \bibinfo {author} {\bibfnamefont {D.}~\bibnamefont
  {Loss}},\ }\href {\doibase https://doi.org/10.1016/j.aop.2011.06.004}
  {\bibfield  {journal} {\bibinfo  {journal} {Annals of Physics}\ }\textbf
  {\bibinfo {volume} {326}},\ \bibinfo {pages} {2793 } (\bibinfo {year}
  {2011})}\BibitemShut {NoStop}%
\end{thebibliography}%
\end{document}